\documentclass{article}

\usepackage{PRIMEarxiv}

\usepackage[utf8]{inputenc} % allow utf-8 input
\usepackage[T1]{fontenc}    % use 8-bit T1 fonts
\usepackage{hyperref}       % hyperlinks
\usepackage{url}            % simple URL typesetting
\usepackage{booktabs}       % professional-quality tables
\usepackage{amsfonts}       % blackboard math symbols
\usepackage{nicefrac}       % compact symbols for 1/2, etc.
\usepackage{microtype}      % microtypography
\usepackage{lipsum}
\usepackage{fancyhdr}       % header
\usepackage{graphicx}       % graphics
\usepackage{amsmath} 
\graphicspath{{media/}}     % organize your images and other figures under media/ folder
\usepackage{cite} % to make citation range [1-4] instead of [1,2,3,4]
\usepackage{float}
\title{Unsupervised denoising of Raman spectra with cycle-consistent generative adversarial networks
}

\author{
Ciaran Bench$^{1}$,  ~Mads Sylvest Bergholt$^{2}$, ~Mohamed Ali al-Badri$^{3,4}$ \\
$^1$Ear Institute, University College London, London, UK~~ \\
$^2$Centre for Craniofacial and Regenerative Biology, Kings College London, London, UK~~\\
$^3$Department of Cell and Developmental Biology, University College London, London, UK~~\\ $^4$Department of Physics, Kings College London, London, UK~~\\ 
\texttt{ciaran.bench@ucl.ac.uk~~mads.bergholt@kcl.ac.uk~~m.al-badri@ucl.ac.uk}\\
}

\begin{document}
\maketitle

\begin{abstract}
Raman spectroscopy can provide insight into the molecular composition of cells and tissue. Consequently, it can be used as a powerful diagnostic tool, e.g. to help identify changes in molecular contents with the onset of disease. But robust information about sample composition may only be recovered with long acquisition times that produce spectra with a high signal to noise ratio. This acts as a bottleneck on experimental workflows, driving a desire for effective spectral denoising techniques. Denoising algorithms based on deep neural networks have been shown superior to `classical' approaches, but require the use of bespoke paired datasets (i.e. spectra acquired from the same set of samples acquired with both long and short acquisition times) that require significant effort to generate. Here, we propose an unsupervised denoising approach that does not require paired data. We cast the problem of spectral denoising as a style transfer task and show how cycle-consistent generative adversarial networks can provide significant performance benefits over classical denoising techniques.
\end{abstract}

% keywords can be removed

\section{Introduction}

The discernment of pathological conditions within tissues based on the molecular constituents of their components represents a fundamental aspect of biomedical research. Pertinent biochemical alterations occurring in the early stages of diseases can be identified through molecular compositional changes \cite{auner2018applications, devitt2018raman,mayne2013resonance,zhang2005rapid,unal2016novel,guerrini2019surface,abramczyk2013raman,smolsky2017surface,tu2012diagnostic}. Raman spectroscopy has emerged as a promising avenue for evaluating the molecular contents of biological specimens, thereby manifesting its considerable potential as a clinical diagnostic tool \cite{lawson1997biomedical,nicolson2021spatially,krafft2015many,choo2002medical}. Its non-destructive nature and ability to be used without contrast agents make it particularly attractive for \textit{in vivo} applications \cite{choo2002medical,bakker2000vivo,buschman2000vivo}. Additionally, its rich chemical information makes it appealing for microscopy and the characterisation of cell phenotypes and tissues \cite{horgan2020molecular,kozik2021review,choo2002medical,kong2015raman}.
However, the effectiveness of Raman spectra in determining molecular composition relies on several factors, including levels of noise in the spectra. To resolve relevant spectral features, long exposure times (1-10 sec) are typically required. Microscopy applications and the generation of datasets for training network-based classifiers often involve the acquisition of hundreds or thousands of spectra, making long acquisition times impractical \cite{palonpon2013molecular, guo2020deep, antonio2014advances,blake2022machine,othman2019reduced,he2021raman,zhang2022raman,yu2019deep,yan2020diverse}. Fibre-optic applications of Raman spectroscopy for \textit{in vivo} diagnosis of diseases (e.g., cancers) is severely hampered by noise and many applications require acquisition times at the order of seconds. Raman spectra generally consist of several noise contributions including signal and background shot noise, detector dark noise and readout noise \cite{mccreery2005raman}. Therefore, efforts have been made to develop efficient spectral denoising techniques capable of resolving important features in spectra acquired more quickly.

\subsection{Supervised spectral denoising}
Classical approaches (and their variants), such as Savitsky-Golay, Wiener filtering, and wavelet denoising can provide adequate improvement to spectral quality in some circumstances \cite{donoho1990does, barton2018algorithm,huang2018fitting,byrne2016spectral,butler2016using,ramos2005noise,liu2015joint,chen2018adaptive,xi2018novel,chen2014recovery, laurent2019denoising,zhao2022adaptive, camp2014high,camp2016quantitative,masia2013quantitative, chen2014recovery,craggs1996maximum}. However, each has its own set of disadvantages \cite{eilers2003perfect,schmid2022and,sharifi2022evaluation,kawala2020comparison,dai2014recent,koziol2018comparison} (e.g. Savitsky-Golay smoothing can struggle to suppress high frequency noise and produce artefacts near the boundaries of data). Recent work has shown that denoising algorithms based on deep neural networks can provide superior performance in a range of applications.

Networks provide a mechanism for learning the optimal operations to perform on a spectrum with a low signal-to-noise ratio (SNR) to restore signal quality. The form of these transformations do not have to be known \textit{a priori} nor do they need to be hand-engineered for a particular dataset, catering to assumptions about how noise emerges in the spectra (made implicitly or otherwise \cite{eilers2003perfect,koziol2018comparison}). In contrast to classical approaches, networks learn the optimal transformations for a particular dataset automatically using feedback about how particular adjustments to their parameters may improve spectral quality. Supervised learning is the the most generic framework where the parameters of these operations are adjusted so as to reduce the difference between low SNR spectra that have been operated on by the network and their corresponding high SNR spectra acquired from the same location/sample. This is executed via some form of gradient descent and backpropagation to compute how parameter changes ultimately effect spectral quality. I.e. given a paired dataset composed of low SNR spectra $x_1, x_2, ..., x_i \in X$ and a corresponding set of spectra acquired from identical samples with higher SNR $y_1, y_2, ..., y_i \in Y$ (where $x_i$ and $y_i$ are acquired from the same location in the same sample), a network $K$ learns to approximate a non-linear function $f$ that ideally transforms an input low SNR spectrum into its high SNR counterpart $f:X \rightarrow Y$, such that $K(x_i)\approx y_i$. 

The most commonly used framework for learned denoising is supervised learning with networks composed of convolutional layers \cite{machado2022deep,pan2020noise,zeng2023modified,loc2022denoising,loc2022raman,yoon2023deep,horgan2021high,Horgan_spectra_ai,zhang2020high,vernuccio2022fingerprint}. Within this framework, convolutional layers detect features spanning the wavenumber dimension of a given low SNR spectrum, and manipulate information about their presence to output a denoised version of the input spectrum. With sufficient downsampling, the receptive field of a fixed-size convolutional kernel implemented in successive convolutional layers will increase, enabling the extraction of features at multiple length scales. Though this comes with an associated loss of resolution along the wavenumber dimension (also affecting information about where a given feature is located along a spectrum). To reduce the negative consequences this may have on the adapted spectral quality (i.e. low resolution output spectra), U-Net like architectures featuring long-range skip connections have been implemented \cite{tang2022two,horgan2021high,Horgan_spectra_ai,gebrekidan2021refinement}. Here, activation maps produced by layers in the encoding pathway (encoding information about simple features at high resolution) are concatenated to those produced by the decoder pathway (containing low resolution information about more complex features that span larger length scales that are composed of ensembles of simpler features). This provides the decoder with high resolution information about where the simple features that are constituents of the more complex features detected downstream by the network lie along the wavenumber dimension, improving the resolution of the outputs.  Horgan et al. and others also report the use of short term ResNet-style skip connections that reduce the effects of model degradation with increasing number of layers, and can help stabilise gradient behaviour \cite{horgan2021high,Horgan_spectra_ai,kung2020baseline,Horgan_spectra_ai}. 

Networks composed of recurrent (RNN) or long-short term memory (LSTM) layers are specialised for extracting and manipulating features from sequential data, retaining context at both long and short ranges. It has been suggested that this framework may confer performance benefits compared to networks based on convolutional layers for classification tasks \cite{yu2021analysis,wang2021discrimination,liu2023spatially}. Though, the practical extent of these benefits compared to convolutional layers is not clear from the literature. Consequently, any potential advantages of using them for spectral denoising is also unclear. This is similarly the case for hybrid architectures composed of both convolutional and LSTM/RNN layers \cite{selvarani2023label,cai2023line}. 

Paired datasets have also been used within generative adversarial learning frameworks (a short summary of generative adversarial networks can be found in \cite{bench2023enhancing}). In \cite{ma2022conditional}, an encoder-decoder style convolutional generator was trained with a convolutional discriminator, where each update to the generator was performed by minimising a loss function featuring a supervised term (i.e. a comparison between an adapted spectrum and its high SNR ground truth counterpart). Given the inclusion of a supervised loss, the benefits of using a generative framework for learning the denoising over a more generic supervised learning scheme is unclear (especially considering no comparison with a purely supervised algorithm was reported). Though, it is possible that using a discriminator may help avoid artefacts in adapted outputs that may occur from a poorly formulated supervised loss function \cite{mustafa2022training}. This approach has also been used in other domains. \cite{al2022generative} used a similar scheme for IR spectra, while \cite{wu2020spectra} trained a cycleGAN-like architecture with a supervised loss term to denoise galactic flux spectra. Conditional GANs have been used for denoising 1D EEG data\cite{wang2022ecg}, though, this particular implementation is essentially supervised given the need to provide the discriminator with paired examples.

\subsection{Unsupervised denoising}

A major drawback with implementing supervised approaches is the need for a paired dataset (i.e. both low and high SNR spectra acquired from from the same location from the same samples) that requires significant effort to generate. This contrasts with unpaired datasets where spectra from either class can be acquired in different locations or from different samples. In principle, it might be possible to use a pretrained denoising network to denoise spectra acquired from some new target sample without any additional training. But this would only provide suitable enhancement if it was known \textit{a priori} whether a given set of test data were described by the same data domain as a pretrained denoising network's training set \cite{blake2022machine}. A data domain $D = \{\chi, \Upsilon, P(x,y)\}$ characterises a given dataset. It consists of an input feature space $\chi$ (a vector space containing all input features), an output feature space $\Upsilon$, and a joint probability distribution $P(x,y)$ over the input and output feature space pair $\chi \times \Upsilon$. Here, $x$ is an instance of the network inputs $x_1, x_2, ... x_i \in X$ and $y$ is an instance of the corresponding ground truths $y_1, y_2, ...y_i \in Y$ \cite{kouw2019review,yang2009heterogeneous, csurka2017domain, pan2009survey, zhao2020review, kouw2018introduction,liu2022deep}. The joint probability distribution can be decomposed into marginal (commonly referred to as the `data distribution') and conditional distributions: $P(x,y) = P(x)P(y|x)$ or $P(x,y)=P(y)P(x|y)$. The generalisability of existing denoising models have not been proven robustly over a significant range of possible spectra (which vary in terms of instrumentation used for acquisition, acquisition settings, and sample type), and it is quite possible that some form of supervised transfer learning (with a small paired dataset from the target domain) would have to be implemented to achieve suitable performance on a given test set with a pretrained network.

It would instead be preferable to use a completely unsupervised framework for learned denoising. Unpaired data is comparatively easier to generate as less care is needed to ensure that the exact same samples/locations are used to produce examples for each domain. Furthermore, with some unsupervised schemes, one could include unpaired examples from preexisting and publicly available high and low SNR spectral datasets. This could significantly reduce the effort required to procure data for training. Though, as will be discussed, the suitability of any publicly available data will depend on the degree of domain alignment it has with any test data one wishes to apply the algorithm to.

Denoising autoencoders are the most widely reported unsupervised denoising techniques \cite{machado2022deep,xu2022high,fan2021signal,brandt2021deep,zeng2023modified}. The network is composed of two modules, an encoder $E$ that learns to approximate a function $f_E$ that maps low SNR input spectra $x_i$ to low-dimensional latent representations $o_i \in O$, so $f_E:X\rightarrow O$, and a decoder module $Q$ that learns to approximate a function $f_Q$ that reconstructs the input spectra from their latent representations $f_Q:O\rightarrow \hat{X}$, where the reconstructed spectra $\hat{x}_1, \hat{x}_2, ..., \hat{x}_i \in \hat{X}$. The network is typically trained end-to-end, minimising an objective function comparing $Q(E(x_i))$ and $x_i$. Provided the compression is strong enough, the latent representation should not have sufficient capacity to encode all of the noise found in the sample, instead prioritising its key structural features. Consequently, reconstructed spectra should be smoother/denoised where ideally $Q(E(x_i))\approx y_i$. One appealing feature of this approach is that it can be trained without any high SNR data. With that said, the operations of both modules are adjusted so as to optimise reconstruction quality; no feedback is used to enforce the accuracy of the smoothing. The fact that the model can enhance spectra in some cases is a useful by-product of using a compressive encoding pathway, rather than a directly optimised feature of the algorithm.

\subsubsection{Generative deep learning for spectral denoising}

Ideally, an unsupervised denoising network should be optimised using an objective function that encodes information about the quality of denoised spectra. One way of doing this would be to assess whether an adapted spectrum appears to belong to the same data distribution describing high SNR spectra. Cycle-consistent generative adversarial networks (cycleGANs) provide a framework for learning the optimal spectral denoising operations based on this kind of feedback \cite{zhu2017unpaired}. CycleGANs have been used to perform completely unsupervised denoising in other domains, such as for 2D seismic data \cite{li2020denoising,li2021residual}, CT image denoising \cite{li2019low}, and digital holographic microscopy image denoising \cite{gupta2023high}. A cycleGAN inspired architecture has also been used for denoising EEG signals in a completely unsupervised manner \cite{gandhi2018denoising}. However, their application to Raman spectroscopic data remains limited, having been used for virtual staining \cite{koivukoski2023unstained,he2022image} or suspension array decoding \cite{yao2022decoding}.

\begin{figure}%[H]
    \centering
    \includegraphics[width=\columnwidth, trim={0 0 0 .3cm},clip]{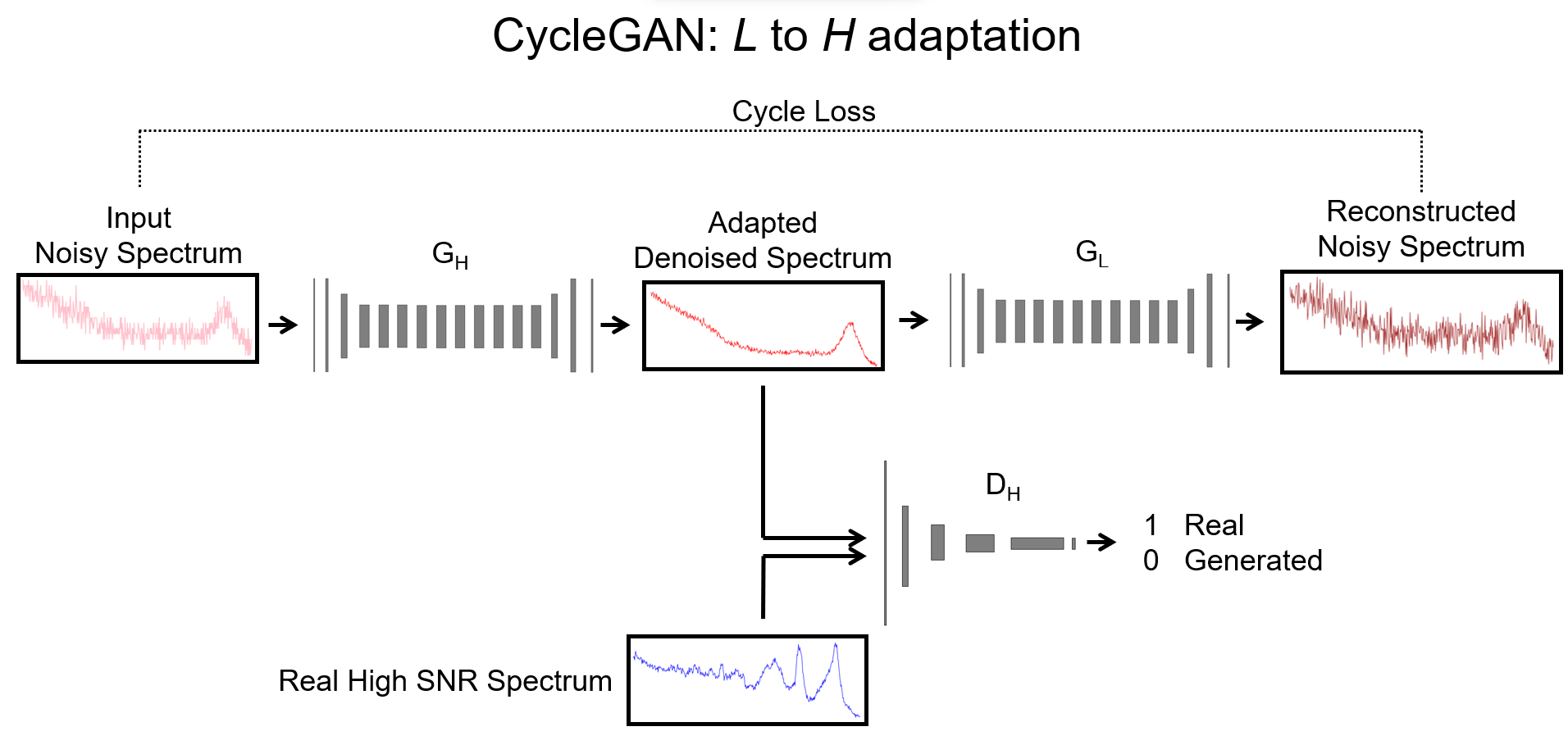}
    \caption{Forward-pass when updating the denoising cycleGAN's $G_H$. A noisy spectrum $l$ is adapted by $G_H$. $G_H(l)$ is then passed through $G_L$, where $l$ and $G_L(G_H(l))$ are used to compute the cycle consistency loss. $l$ and $G_L(l)$ (not shown here) are also used to compute the identity loss. Both real high SNR spectra, and those generated by $G_H$ are classified by $D_H$. Unpaired spectra are used to train the network. For the forward pass shown here, a water spectrum is adapted by the generators, while the discriminator is fed a high SNR spectrum of cell proteins. All of the steps involved with training are described in Appendix \ref{sec:cycleGAN_train}.}
    \label{fig:LH_adaptation}
\end{figure}

CycleGANs are composed of four networks, two generators ($G_H$ and $G_L$) and two discriminators ($D_H$ and $D_L$). $G_H$ will take a sample from a dataset $L$, described by a feature space $\chi_{L}$ and data distribution $p(l)$, where examples $l_1, l_2, ... l_i \in L$, and $l$ is an instance of this dataset and adapt it so it would appear to belong to dataset $H$, described by a feature space $\chi_{H}$ and a data distribution $p(h)$, where examples $h_1, h_2, ... h_i \in H$, and $h$ is an instance of the dataset. Examples from $L$ and $H$ are unpaired. $G_L$ will adapt a spectrum (e.g. a spectrum from $H$), so it would appear to belong to $L$. $D_L$ learns to differentiate real examples from $L$ and those produced by $G_L$. $D_H$ takes real examples from $H$, and those generated by $G_H$, and learns to differentiate the two. A cycleGAN is trained using loss terms designed to constrain the adaptations, ensuring that an adapted example still retains much of its original structure. This is achieved through the use of a cycle consistency loss (e.g. a comparison between $G_H(G_L(h))$ and $h$ as well as $G_L(G_H(l))$ and $l$). An identity loss, (a comparison between $G_L(l)$ and $l$ as well as $G_H(h)$ and $h$) is used to ensure interchannel features are preserved. In our case, $L$ consists of low SNR spectra, while $H$ consists of high SNR spectra (see Fig. \ref{fig:LH_adaptation}). The adaptation learned by $G_H$ should be directly optimised to ensure that transformed spectra appear to belong to $H$. This form of feedback, though less direct than that provided with supervised learning, could still provide superior performance compared to classical and autoencoder based denoising strategies that do not have such feedback mechanisms.

Any test data provided to $G_H$ should ideally be drawn from $p(l)$. Any high SNR spectra belonging to $H$ should have the desired properties one wishes to see in the adapted test spectra (e.g. resolution, SNR). Ideally, any test data and training data should consist of unpaired example spectra acquired from the same kinds of samples and using the same instrumentation. This will help ensure that adaptations do not add spectral features that would not normally occur in the spectra acquired from the test samples. Though, the constraints imposed by the cycleGAN minimise the possibility of this even when there is mismatch in the feature spaces of the test data and $H$.

It is worth noting that, aside from the cycleGAN, variants of all the schemes mentioned here have also been implemented for spectroscopic hyperspectral image denoising \cite{abdolghader2021unsupervised,manifold2019denoising,lin2019deep,yuan2018hyperspectral,xie2018deep,he2019speeding}. Though our focus in this article remains on the adaptation of singular spectra.

\subsection{Objectives}
Here, we train a cycleGAN to denoise Raman spectra acquired from confocal Raman hyperspectral images of MDA-MB-231 breast cancer cells. Our training and test sets were the same as that in used in \cite{horgan2021high}, but reorganised so training could be conducted in a completely unsupervised framework.

One key challenge with training cycleGANs is formulating a fully unsupervised and robust stopping criteria \cite{nerrienet2023standardized}. Such a metric should encode information about the degree of enhancement the network has applied to the spectra. Generic signal quality metrics (e.g. SNR) are fundamentally insufficient as, in addition to enhancing signal quality, we are also interested in ensuring that key spectral features present in the input spectra are reconstructed accurately. To this end, we propose a stopping criterion based on the quality of the cluster groups formed by fitting adapted spectra (adapted low SNR validation spectra) to the clusters derived from a set of high SNR validation spectra (unpaired with the adapted spectra). Provided both sets of validation spectra are acquired from the same types of samples, then appropriately denoised low SNR validation spectra should be well classified by the cluster groups fit to the high SNR validation spectra. We validate the efficacy of this unsupervised metric by showing it correlates strongly with a traditional supervised loss.

To compare with another unsupervised method, we also train a denoising autoencoder. We show how a cycleGAN can indeed outperform both classical and autoencoder-based denoising techniques. Ultimately, this framework provides a means to significantly improve experimental throughput without the need to procure a paired dataset. We anticipate this will reduce the effort required to use highly effective network-based data processing techniques into experimental workflows. 

\section{Methods}
\subsection{Data preparation}
We trained the denoising cycleGAN on an adapted version of the MDA-MB-231 breast cancer cell spectral dataset used in \cite{horgan2021high}. The unadapted training set was composed of low SNR (0.1s integration time per spectrum) and high SNR (1s integration time per spectrum) hyperspectral cell image pairs of varying size (consisting of a total of 11 cells) using 532 nm laser excitation. These spectra carry molecular information about lipids, proteins, and nucleic acids, and consequently provide a robust challenge for the network in terms of the varied chemical information it must learn to preserve with the adaptation. Furthermore, being composed of low and high SNR spectra acquired from the same kinds of samples, this dataset allows us to test the algorithm in an ideal case where there is likely to be a significant degree of data domain alignment between examples from the high and low SNR domains. Consequently, this allows us to assess an upperbound on the performance we could expect from this approach when applied to data acquired from real biological samples. The data was pre-augmented with spectral shifting, flipping, and background subtraction (as described in \cite{horgan2021high}). The unsupervised/unpaired training set was constructed by first randomising the order of the data pairs (while preserving each individual data pair). The first $80\%$ of spectra were set aside for the training set, while the remaining $20\%$ were set aside for the validation set. The training set was constructed by parsing the low SNR spectra from the first half of the training data pairs along with the high SNR spectra from the remaining training data pairs. This resulted in a training set composed of 63847 high SNR spectra, and 63847 low SNR spectra.  The same protocol was used to parse examples from the validation data pairs (composed of 15962 high SNR spectra, and 15962 low SNR spectra). Consequently, both training and validation sets contained only unpaired examples. A version of the validation set with low/high SNR data pairs preserved was also produced to validate our unsupervised sopping criteria, but was not used to facilitate training. The original test set used in \cite{horgan2021high} (12694 paired examples) was also used as our test set without any modifications. No additional normalisation/standardisation was performed on any of the data used as an input to the network.

\subsection{CycleGAN architecture and training details} 
The architectures used for each module of the cycleGAN are shown in Fig. \ref{fig:cycleGAN_archs} in Appendix \ref{sec:cycleGAN_train}. All networks used in this study were written to run with TensorFlow 2.12.0. The code was adapted from \cite{Nain2020}. The procedure for each training batch is described in Appendix \ref{sec:cycleGAN_train}, which includes information about loss functions and hyperparameters (Table \ref{tab:hyperparams}).
\begin{figure}%[H]
    \centering
    \includegraphics[width=.7\columnwidth]{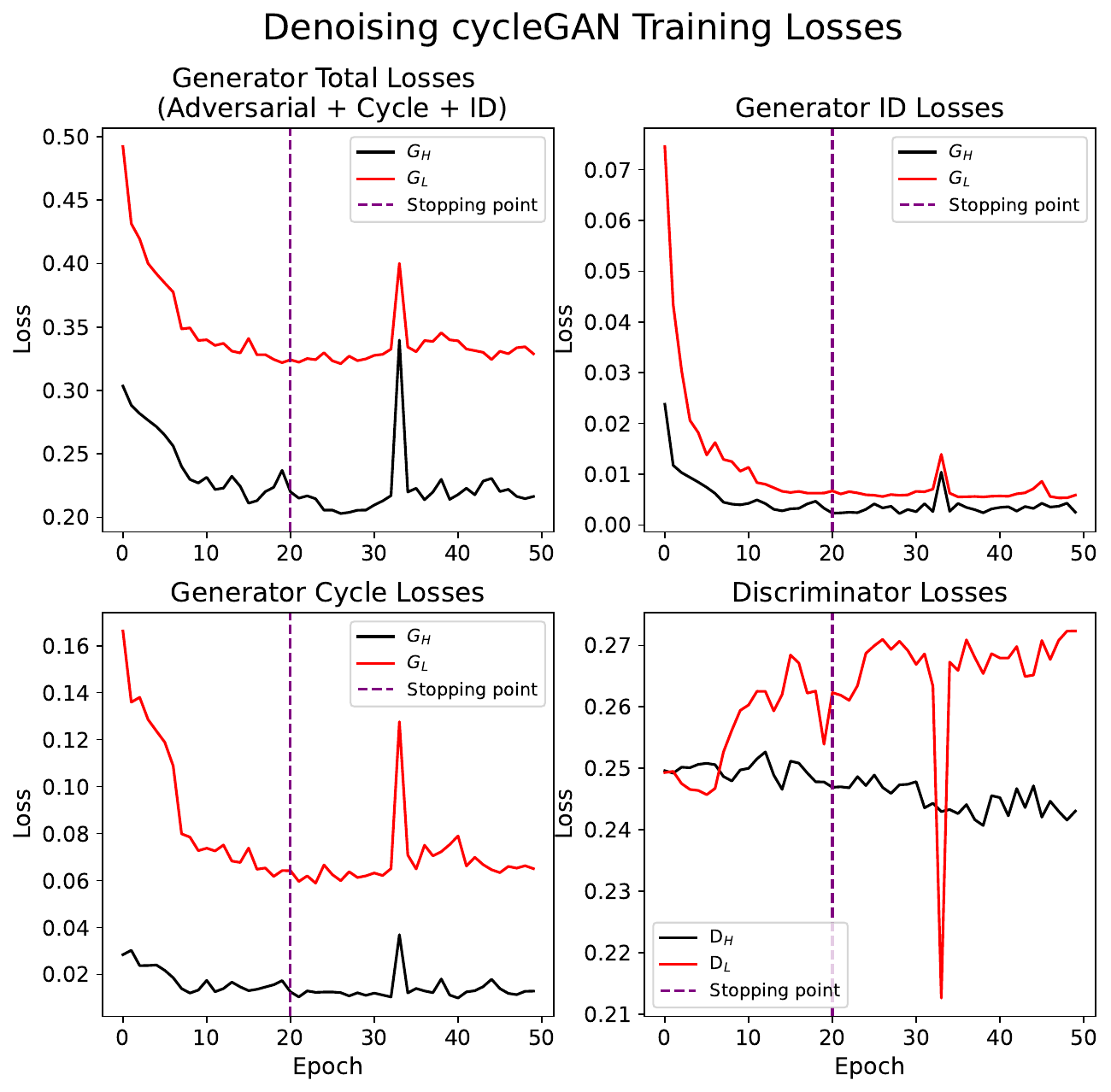}
    \caption{CycleGAN training losses. The purple line shows the epoch used to produce the results described in this article. This stopping point corresponded to the point where the unsupervised validation loss began to plateau (see Fig. \ref{fig:cycleGAN_validation_losses}).}
    \label{fig:cycleGAN_training_losses}
\end{figure}

\begin{figure}%[H]
    \centering
    \includegraphics[width=.75\columnwidth]{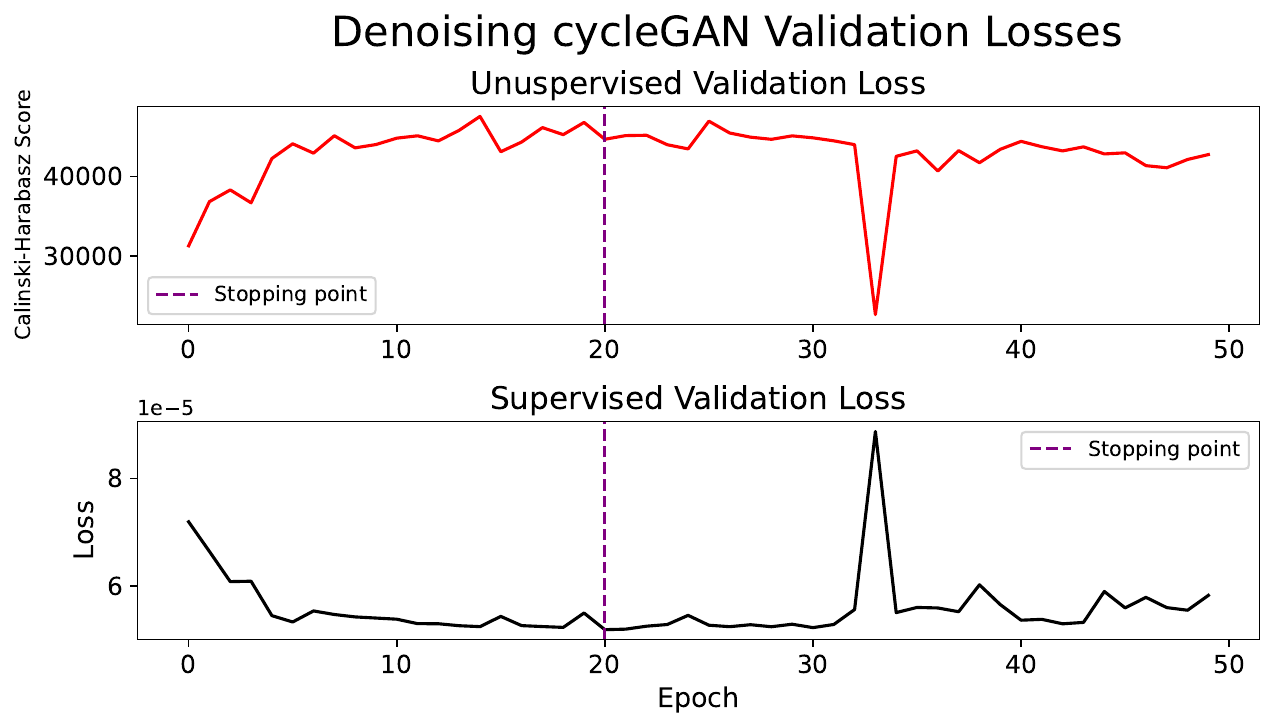}
    \caption{Unsupervised validation loss used to determine the stopping point (purple dashed line), along with a supervised validation loss for the same set of examples used to assess the efficacy of the unsupervised loss. The Pearson's correlation coefficient between the two losses was -0.96, indicating that the unsupervised loss could be used as a suitable proxy for a traditional supervised validation loss.}
    \label{fig:cycleGAN_validation_losses}
\end{figure}

The training loss curves are shown in Fig. \ref{fig:cycleGAN_training_losses}. The validation loss and stopping criteria were calculated as follows. First, the high SNR validation spectra were clustered using the k-means++ algorithm (scikit-learn) with eight clusters (see Fig. \ref{fig:clust_quality_valid_set} in Appendix \ref{sec:cluster_details}). This number of cluster groups was chosen as it was found to provide the maximum Calinski-Harabasz Score (scikit-learn) for this set of spectra, meaning the resulting clusters were the most well formed with the intercluster dispersion mean being maximally higher relative to the intracluster dispersion \cite{calinski1974dendrite}. Therefore, these cluster groups are not optimised to distinguish spectra by whichever tissue subtype they were acquired from (as is traditionally the case with using clustering techniques on spectra). This is because some spectra have been augmented and may not reflect sample composition accurately (and instead, provide a more varied set of features for the network to learn from). Even if the assigned classes are not guaranteed to be tied to a spectrum's corresponding sample type/region, they nevertheless provide us a means to assess the performance of the denoising algorithm over different spectral types found in the test set as defined by their unique spectral features. As will be shown, this is sufficient for formulating a suitable stopping criteria.

Each cluster centre and the number of validation spectra assigned to each cluster is given in Fig. \ref{fig:clust_valid} and Table \ref{tab:clust_valid} in Appendix \ref{sec:cluster_details}. After each epoch, the low SNR spectra were adapted to appear as though they were high SNR, and subsequently clustered according to the cluster groups derived from the high SNR validation spectra. The Calinski-Harabasz Score was then calculated on the clustered denoised spectra. The model state corresponding to the epoch before the Calinski-Harabasz Score began to decrease or plateau (determined by visual inspection) was treated as the trained model. Because the high and low SNR spectra used for training are recorded from the same types of samples, if denoising performance is adequate and spectral features are effectively preserved, then the resulting spectra should appear as though they have been drawn from the same data distribution describing the high SNR validation spectra and consequently produce well-formed clusters (i.e. high intercluster dispersion mean relative to the intracluster dispersion) around the centres derived from the high SNR validation spectra. We validate the efficacy of this unsupervised validation metric by comparing it with a traditional supervised loss (i.e. mean over the mean squared error calculated for each denoised spectrum), and show the two correlate strongly and provide the same stopping point (Pearson's correlation coefficient $c = -0.96$, see Fig. \ref{fig:cycleGAN_validation_losses}). The network's weights at epoch 20 were the most optimal according to the stopping criteria. All validation losses reported in this study were calculated from spectra that had not been post-processed (i.e. baseline corrected) or normalised. This is in contrast to the test error metrics which were computed using baseline corrected and normalised spectra. This accounts for the difference in magnitude observed in the reported supervised validation and test error metrics. 

\subsection{Autoencoder denoising}
\begin{figure}%[H]
    \centering
    \includegraphics[width=.75\columnwidth]{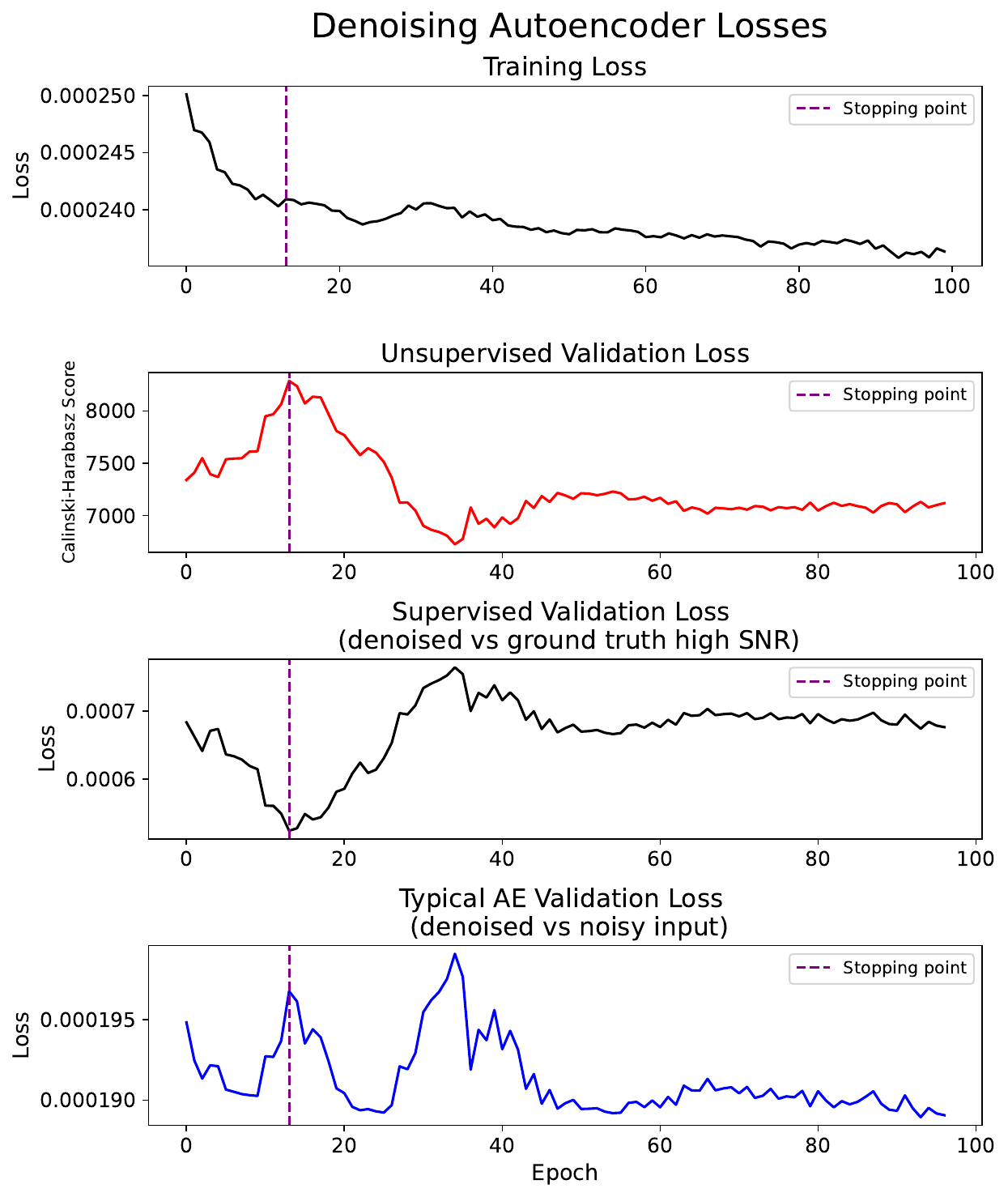}
    \caption{Autoencoder training and validation losses. The dashed purple line shows the epoch used to produce the results described in this article. This stopping point corresponded to the point where the unsupervised validation loss plateaued. Both supervised validation losses are the mean squared error.}
    \label{fig:AE_validation_losses}
\end{figure}

A denoising autoencoder was trained on the same set of low SNR spectra used to train the cycleGAN. The network architecture is shown in Fig. \ref{fig:AE_arch}, with training hyperparameters described in Table \ref{tab:AE_hyperparams} in Appendix \ref{sec:AE_details}. The loss curves are shown in Fig. \ref{fig:AE_validation_losses}. The network was trained using a mean squared error loss function. The structure of the architecture was inspired by other examples in the literature, and the layer parameters were found to produce more optimal performance compared to other variations. Though, an exhaustive parameter/architecture search was not conducted. To provide an upper bound for the performance of the model, the stopping point was determined using a supervised loss (i.e. the same supervised validation loss computed in the cycleGAN study). The same unsupervised loss metric used to train the cycleGAN was also calculated and was similarly found to correlate strongly with the supervised loss with a Pearson's correlation coefficient of $c = -0.99$ (see Fig. \ref{fig:AE_validation_losses}). This provides further evidence of its robustness as an indicator of spectral quality. The network's weights at epoch 13 were the most optimal according to the stopping criteria.

\subsection{Classical denoising approaches}
\label{sec:classical_method}
We compare the performance of the denoising cycleGAN and autoencoder to three commonly used classical denoising approaches: Savitsky-Golay smoothing (scipy.signal.savgol\verb|_|filter), Wiener filtering (scipy.signal.wiener), and wavelet denoising (skimage.restoration.denoise\verb|_|wavelet). We perform each classical denoising technique over a range of parameterisations on the test spectra without any additional pre-processing applied to them. Savitsky-Golay smoothing was performed using the scipy.signal.savgol function with a range of different window lengths, with order set to 3, and \verb|mode="nearest"|. Similarly, wavelet denoising was performed by applying the skimage.restoration.denoise\verb|_|wavelet function over a range of wavelet levels using the Bayes shrink method on soft mode with \verb|sym8| wavelets with \verb|rescale_sigma| set to `True'. Only up to nine levels were considered, as performance appeared to saturate at low numbers of levels. Each instance of Wiener denoising was performed using the scipy.signal.wiener function with default parameters. After denoising, spectra were baseline corrected using the BaselineRemoval python package, with the degree set to three, and then normalised by their resulting maximum value. The ground truth test spectra were also baseline corrected and normalised in the same manner. These post-processed spectra were used for the error calculations. All denoised spectra and ground truth high SNR spectra mentioned in this study were post-processed this way for making comparisons.

In principle, the mean over all of the mean squared errors computed for each spectral pair in the test set (referred to here as the `total mean over the mean squared errors' \textbf{TMMSE}) could be used to derive the optimal parameterisations for each classical technique. However, this metric is biased towards any dominant class of spectrum that may be present in the test set. Consequently, it could be that the parameterisation associated with the lowest TMMSE is the one that is only the most effective for denoising spectra in this dominant class as opposed to the one that is more equally effective across all spectral classes. Instead, it would be preferable to compute a mean error for all the spectra in a given class, and find the parameterisation that produces the lowest mean over these class-specific errors, which we refer to as `mean over class-specific errors' or \textbf{MCSE}. I.e. MCSE $= \frac{1}{C}\sum^{C}_{c=1}\text{MMSE}_c$, where $\text{MMSE}_c = \sum^{S_c}_{i=1}(G_H({l_i})-h_i)^{2}$ is the mean over the mean squared error for each spectrum in class $c$, where $S_c$ is the total number of spectra in class $c$, $i$ indexes each spectrum in a class, $G_H({l_i})$ is a denoised spectrum, and $h_i$ is the corresponding ground truth), and $C$ is the total number of classes. However, the MCSE is still not guaranteed to be completely invariant to class imbalance, as it will be heavily skewed towards any class that a given denoising technique performs especially well (or poorly). For example, in the case of the cycleGAN or autoencoder, this could be the class most commonly represented in the training set. To help account for this, (and to assess whether this particular bias may be present) we also report the standard deviation of the MCSE for each denoising technique along with the standard deviation of the TMMSE. All of these metrics provide a comprehensive overview of the network's performance over the test set.

The MCSE can only be calculated if the class of each spectrum is known \textit{a priori}. This information is not available in our case. Instead, we infer the class of a given spectrum by i) clustering the high SNR ground truths of the test dataset, and ii) assigning the same cluster ID to their adapted counterparts; this makes it possible to calculate the mean of the mean squared errors (MSEs) associated with all the spectra in a given inferred cluster, which would be akin to calculating the mean of the MSEs of all the spectra in a particular class. 

To perform this, we computed the ideal number of cluster groups for clustering the high SNR ground truth spectra using the same procedure used to derive the ideal number of clusters for computing the unsupervised validation loss. This also resulted in a cluster number of eight. Each cluster centre and the number of test spectra assigned to each cluster is given in Fig. \ref{fig:clust_test} and Table \ref{tab:clust_test} in Appendix \ref{sec:cluster_details}. 

\begin{figure}
    \centering
    \includegraphics[width=.7\columnwidth]{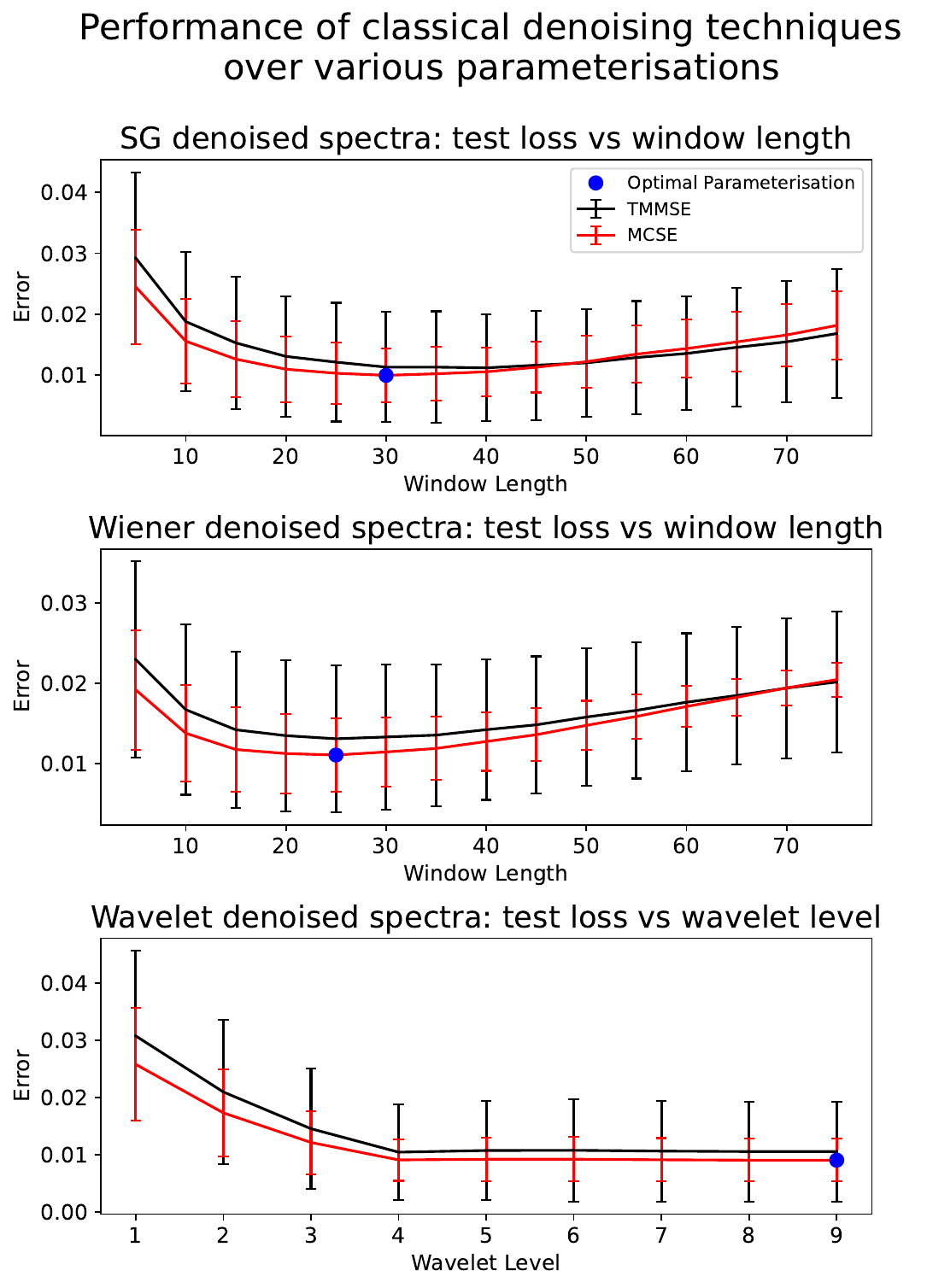}
    \caption{The TMMSE and MCSE computed for each classical denoising technique. The optimal parameterisation for each classical denoising technique corresponded to that which produced the lowest MCSE score (indicating the most effective denoising performance over all inferred spectral classes).}
    \label{fig:classical_params}
\end{figure}

\subsection{Hyperspectral image evaluation}
To assess how well the cycleGAN preserves molecular information, we denoised all of the spectra contained in a hyperspectral image (HSI) of a MDA-MB-231 breast cancer cell (acquired using the same parameters/equipment as the training/validation/test sets), and analysed spectra from regions known to contain either cytoplasm or nucleic acids (determined from the analysis in \cite{horgan2021high}). Given the parameters used to perform the normalisation of the training, validation, and test sets were unknown (this information was not reported with the dataset used in this article), the spectra from this test image were normalised in an ad-hoc manner to roughly match the range of the training set. Before denoising, each spectrum was processed by subtracting their respective minima and then dividing by half of the maximimum value over all spectra in the image. No other processing was applied to the spectra before denoisinig (e.g. cosmic ray correction). After denoising, each spectrum was baseline corrected and normalised following the same procedure as described in Section \ref{sec:classical_method}. The same post processing procedures were applied to the ground truth spectra to enable a comparison (similarly, no additional processing, such as cosmic ray correction, was carried out on these spectra). The low SNR images shown in Figs. \ref{fig:eval_HSI} and \ref{fig:eval_HSI_extended1} exhibit image data that had only been post-processed (only low SNR spectra used for the cycleGAN and wavelet denoising had the pre-processing applied). Because of the ad-hoc normalisation of the input low SNR spectra, it is unlikely that the performance of the network is as optimal as it could be. Nevertheless, the TMMSE of the cycleGAN denoised HSI spectra (TMMSE$ = 0.0044$) was lower than that achieved with spectra denoised with the best performing classical technique (wavelet denoising parameterised with 9 levels had a TMMSE$ = 0.0082$). Furthermore, the denoised HSI spectra appeared qualitatively similar to the denoised test spectra (i.e. no additional artefacts appeared to be introduced) suggesting this normalisation was sufficient for achieving adequate performance with the cycleGAN.

\section{Results and discussion}
\begin{figure}%[H]
    \centering
    \includegraphics[width=.8\columnwidth]{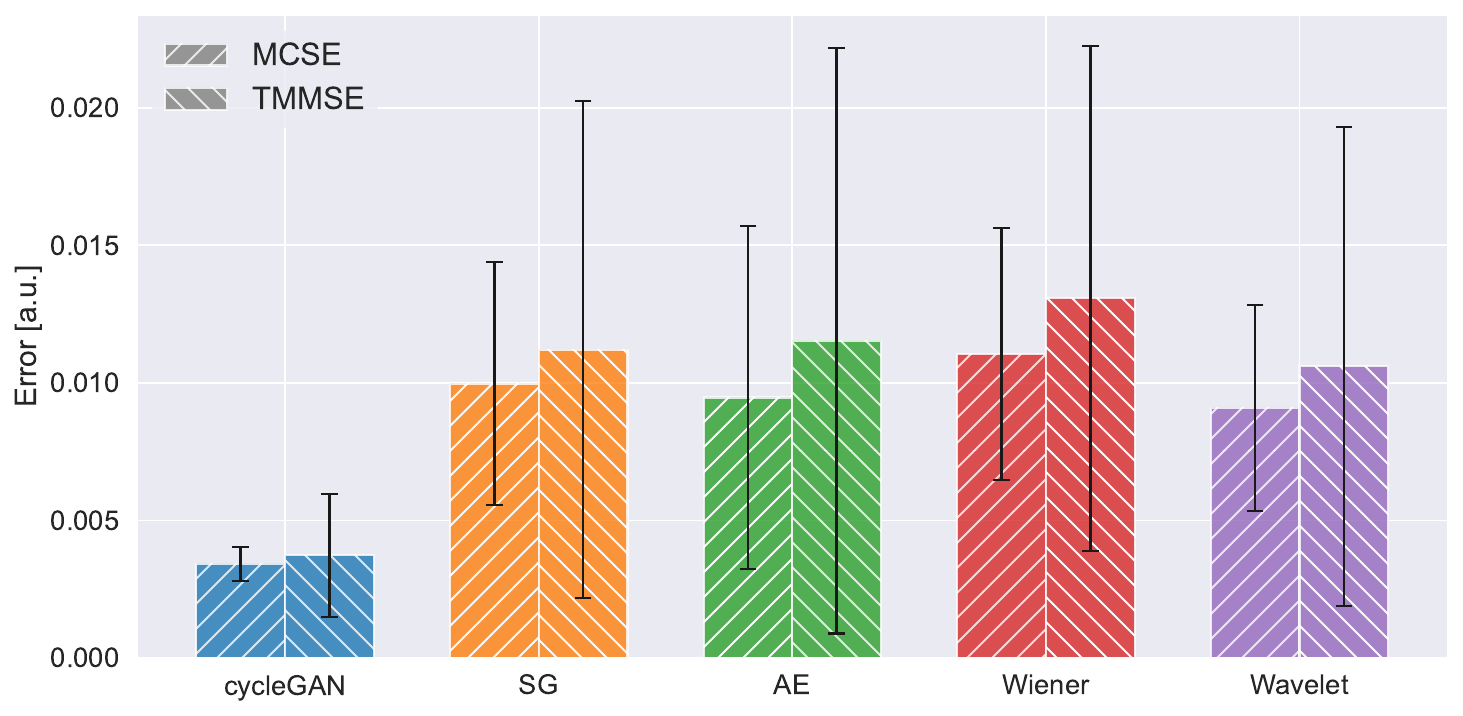}
    \caption{TMMSE and MCSE for all denoising techniques (the metrics for the optimally parameterised classical approaches are shown). The results for the Wavelet technique parameterised with 9 levels, the Savitsky-Golay technique parameterised with a window length of 30, and Wiener filtering parameterised with a window length of 25 are shown. The cycleGAN performed the best compared to all other techniques, and had the lowest variance for both metrics.}
    \label{fig:MSE_test}
\end{figure}

Fig. \ref{fig:classical_params} shows the performance of the non-network based denoising approaches over various parameterisations. Fig. \ref{fig:MSE_test} shows the TMMSE and MCSE for all denoising techniques performed with their their optimal parameterisations (those that produced the lowest MCSE). These metrics are also shown in in Table \ref{tab:test_acc_inferred_classes}.
\begin{figure}
    \centering
    \includegraphics[width=.95\columnwidth]{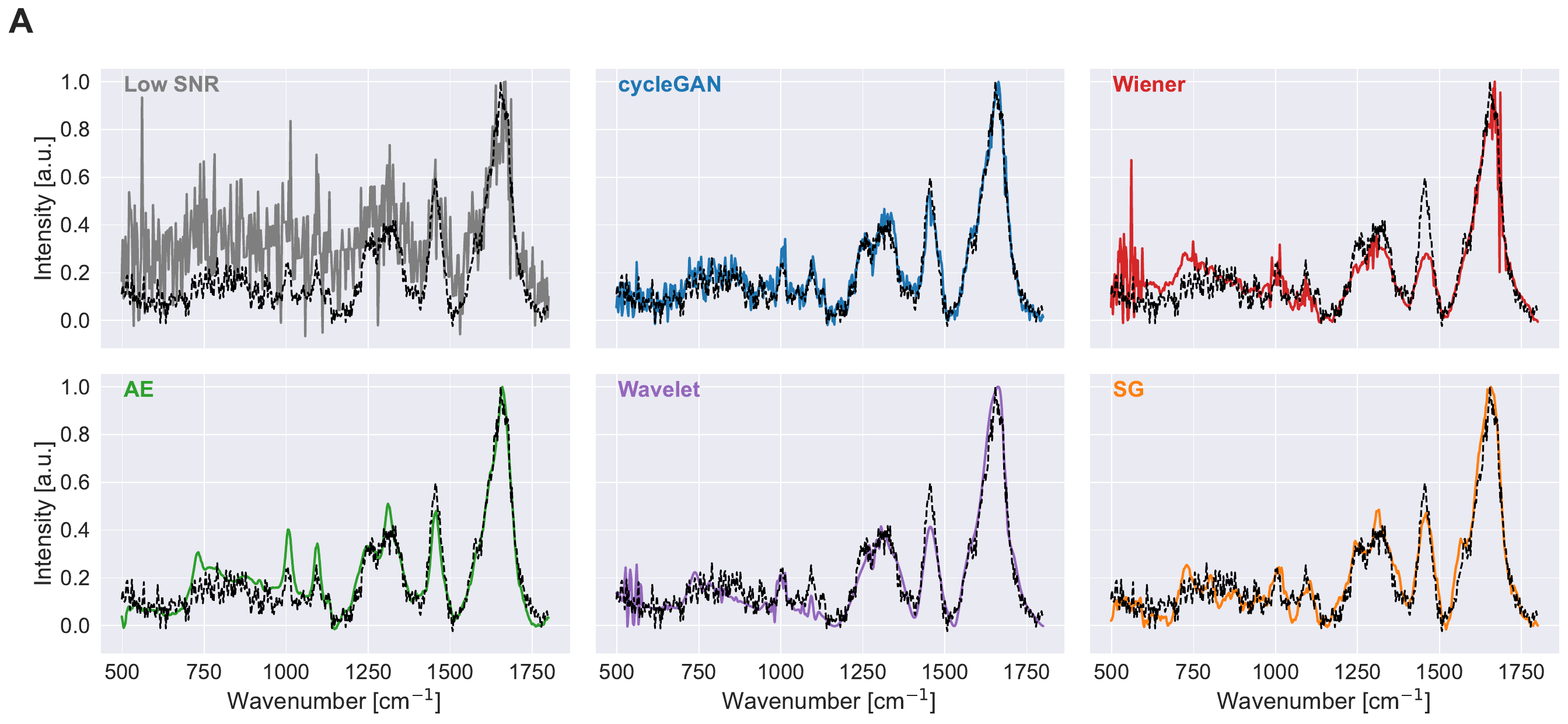}
    \includegraphics[width=.95\columnwidth]{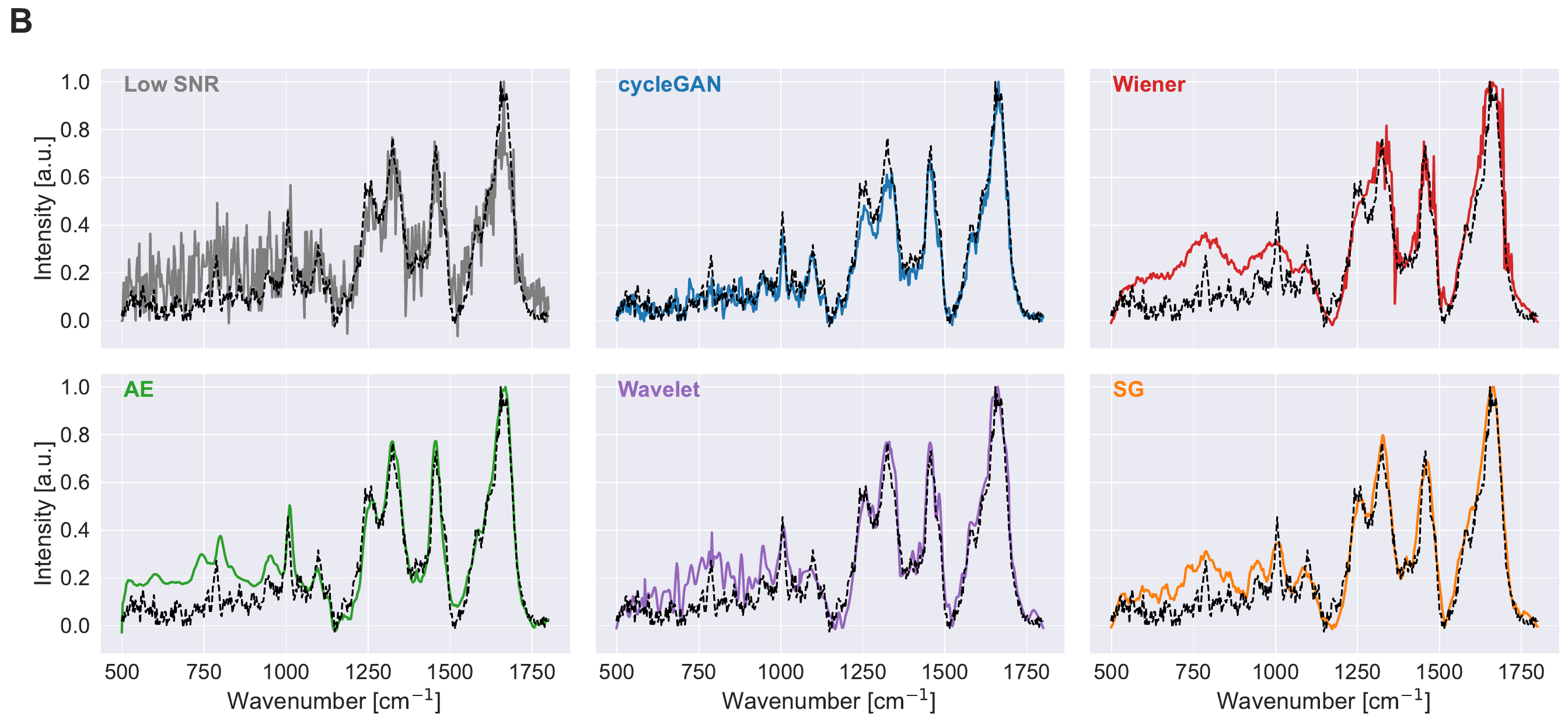}
    \includegraphics[width=.95\columnwidth]{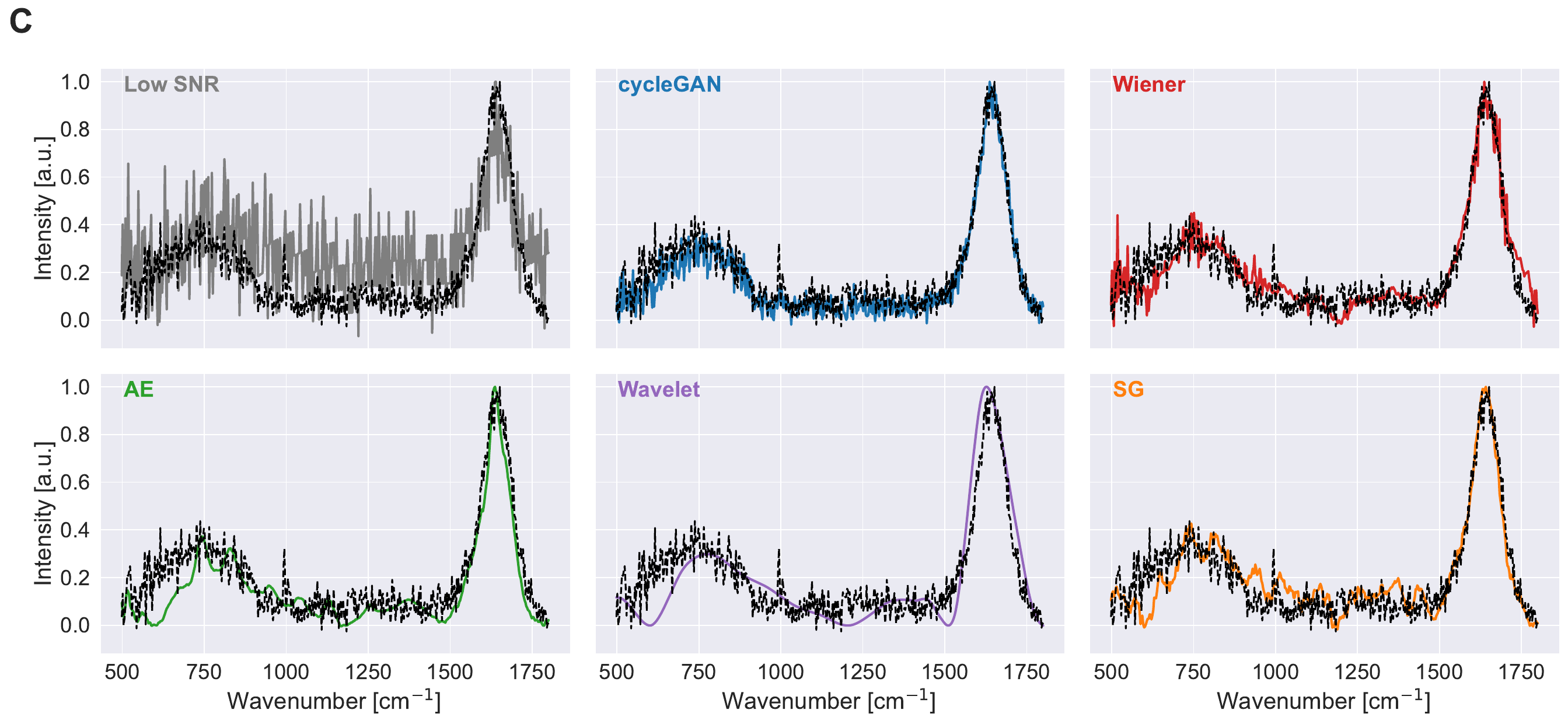}
    \caption{\textbf{A)}, \textbf{B)}, and \textbf{C)} show the performance of each denoising technique on a particular test spectrum (black dotted line is the high SNR ground truth). Spectra were denoised without any preprocessing. For plotting (as shown here) and computing the error metrics, the denoised spectra were baseline corrected and then normalised by their max values.}
    \label{fig:denoised_test}
\end{figure}
The cycleGAN provided clear performance benefits compared to all other denoising approaches. The smaller variance in the cycleGAN's TMMSE and MSCE indicate that its performance is much more consistent over the whole range of test spectra that were evaluated. The superior performance of the cycleGAN over the denoising autoencoder suggests that optimising parameters using an objective function encoding information about spectral quality is key to achieving optimal performance. Looping over the test set with a batch size of 200 for evaluation, the whole test set was evaluated and saved to a file in 2.43 seconds using an Nvidia GeForce 4090 GPU on a local workstation. This demonstrates that the algorithm has clear potential to increase experimental throughput.

It is important to note that the cluster group assigned to a given spectrum in the MCSE calculation may not strictly correspond to the type of sample it was acquired from, as is usually the aim with performing cluster analyses on spectral data. This is because some spectra have been augmented, and may no longer reflect sample composition accurately (and instead, provide a more varied set of features for the network to learn from). Even if the assigned classes are not guaranteed to be tied to a spectrum's corresponding sample type/region, they nevertheless provide us a means to assess the performance of the denoising algorithm over different spectral types found in the test set as defined by their unique spectral features. It is of interest to note that the unsupervised validation loss used for training the cycleGAN is likely less susceptible to biases from class imbalances, as it is calculated based on inter and intracluster dispersion. Instead, it may be biased towards clusters with outlying dispersion properties that may skew the Calinski-Harabasz Score. The fact that the network had comparable performance on all inferred classes in the test set indicates that if any such biases were present, they did not negatively impact the efficacy of the stopping criteria in a significant way. 

\begin{table}[H]
 
  \centering
  \caption{TMMSE and MCSE for each denoising technique.}
  \begin{tabular} {ccccc}
  %\begin{tabular}{lllll}
    \toprule
    \multicolumn{5}{c}{Test accuracy}                   \\
    \cmidrule(r){1-5}
    Technique     & TMMSE & $\sigma$ & MCSE & $\sigma$     \\
    \midrule
    cycleGAN &0.00372 & 0.00224& 0.00341 & 0.00061\\
    SG & 0.0113  &0.00902& 0.00997  &  0.00443  \\
    Autoencoder  & 0.0115& 0.0106&0.00946 &0.00622 \\
    Wiener  &0.0131 & 0.00918&0.0110  &0.00459  \\
    Wavelet      &0.0106  &0.00870& 0.00909 & 0.00373 \\
    \bottomrule
  \end{tabular}
  \label{tab:test_acc_inferred_classes}
\end{table}

Imaging of the characteristic Raman peaks associated with phenylalanine of proteins ($\sim$ 1005 cm$^{-1}$), nucleic acids ($\sim$ 1336 cm$^{-1}$) and lipids ($\sim$ 1450 cm$^{-1}$, CH$_2$ deformations of lipids) in the denoised test HSI are shown in Figs. \ref{fig:eval_HSI} in the main text and S6 in the supplemental document. Despite the network's lack of spatial context, the quality of the images improved considerably compared to their low SNR counterparts. Spectra acquired from the red and white pixel regions (cytoplasm and nucleic acids respectively) are shown in Fig. \ref{fig:eval_HSI}, where the cycleGAN appeared to preserve the molecular information contained in the spectra more accurately than the wavelet denoising.

It is expected that a network using spatial context could achieve superior image denoising performance than than that shown in Fig. \ref{fig:eval_HSI} \cite{koziol2018comparison,yuan2018hyperspectral}. However, given the high dimensionality of HSI data, training networks that utilise spatial context is non trivial in terms of accommodating the large memory requirements and curating a large enough dataset to reduce the ill-effects of the curse of dimensionality. In any case, our primary aim of showing the results in Fig. \ref{fig:eval_HSI} was not to demonstrate our 1D cycleGAN's ability to denoise images, but rather to show where the spectra shown in Fig. \ref{fig:eval_HSI} were located within the cell. 

\begin{figure}
    {\centering
    \includegraphics[width=\columnwidth]{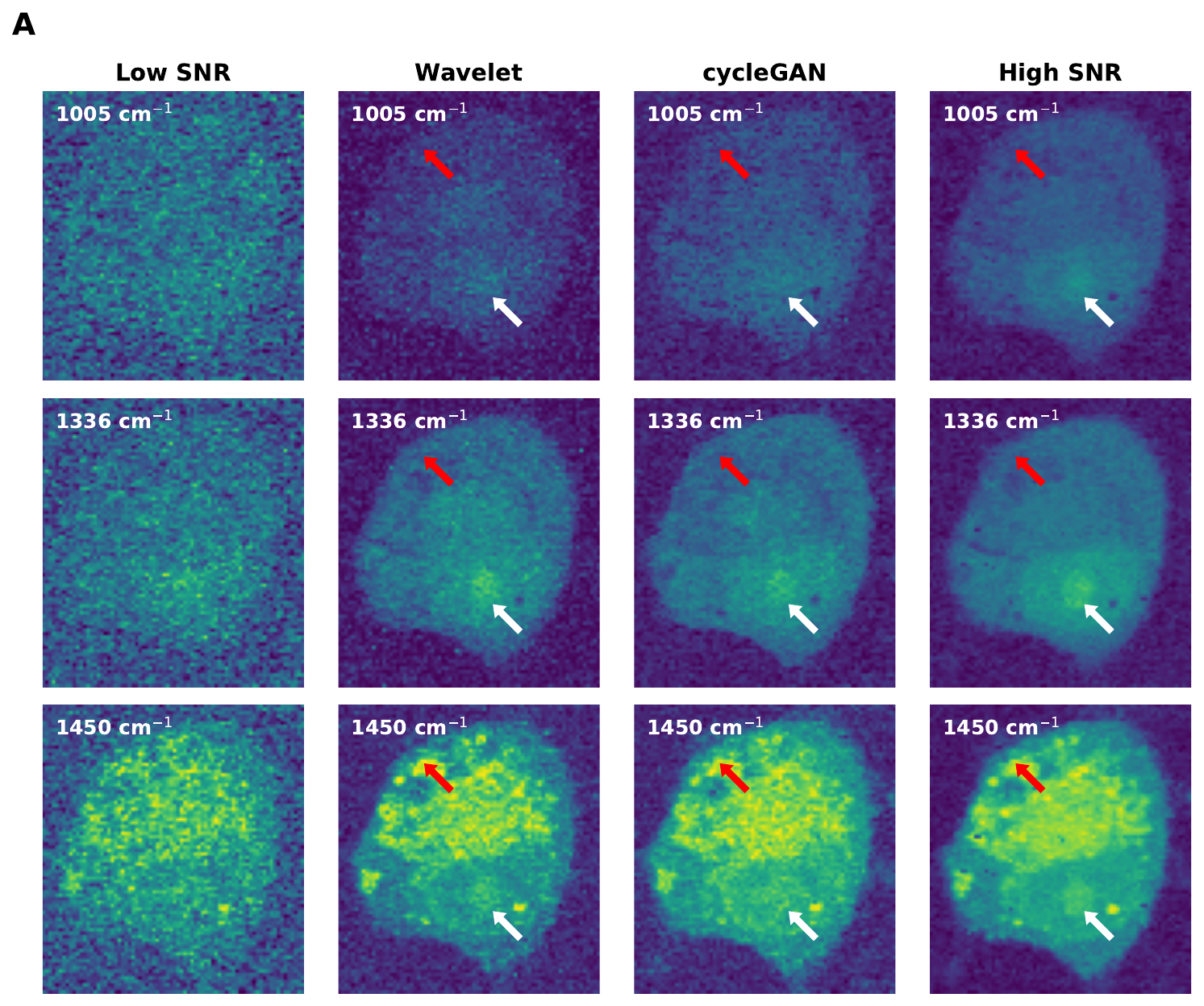}
    }
    \includegraphics[width=\columnwidth]{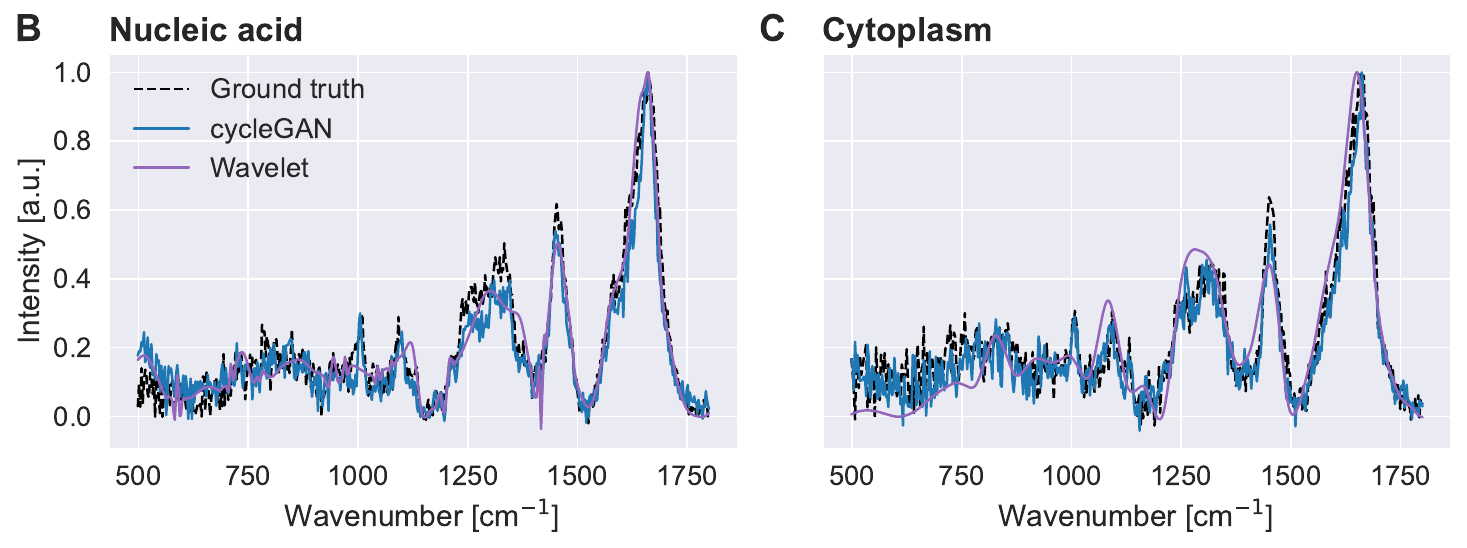}
    \caption{\textbf{A)} 2D slices of the test HSI before and after denoising (where the wavelet denoising was parameterised with 9 levels), as well as the ground truth high SNR slices for 1005 cm$^{-1}$ (typically associated with the phenylalanine in proteins), 1336 cm$^{-1}$ (typically associated with nucleic acids), and 1450 cm$^{-1}$ (typically associated with lipids). Each image has a height of 41.5 $\mu m$, and width of 37.5 $\mu m$. Each spectrum was normalised independently of the others. The red and white arrows show regions with high lipid (cytoplasmic) and nucleic acid content respectively (as determined from the analysis in \cite{horgan2021high}). \textbf{B)} spectra from the white (nucleic acid) and \textbf{C)} red (cytoplasm) regions of the cell. The cycleGAN has better preserved the form of the spectra compared to the wavelet denoising technique.}
    \label{fig:eval_HSI}
\end{figure}

It is evident that the network has performed well on the test data. However, this study represents an ideal case where the spectra from both the high and low SNR domains were acquired from the same kinds of samples. In practice this may not be feasible, and so it remains to be seen how effective this approach may be when the adaptation is learned between spectra acquired from different sample types. The accuracy reported here could be treated as an upper bound estimate for how well this technique may work on real data. Further experimentation with architecture hyperparameters or with the use of bespoke loss functions that are more effective at capturing sparse features in spectra \cite{barton2021convolution} have the potential to confer performance benefits. However, neither of these were investigated here. It also remains unclear as to whether the performance of this network may generalise to spectra acquired from other systems, sample types, or acquired over different wavenumber regions. It might be possible to use transfer learning to adapt its performance in cases where even unpaired examples are scarce and consequently when it is not possible to train a denoising network `from scratch'. Such strategies have been implemented for supervised denoising networks \cite{horgan2021high,Horgan_spectra_ai}. Additionally, this work only compares the performance of the cycleGAN to a subset of denoising techniques, and only one version of the denoising autoencoder (it is possible that other architectures may provide superior enhancement to spectra). Therefore, the full extent of any performance benefits it may incur compared to other approaches remains unclear \cite{guo2022iterative}. Nevertheless, we have shown that the cycleGAN framework can clearly provide performance benefits over the other techniques that were described here. We hypothesis that this is because its parameters are specifically optimised using an objective function encoding information about spectral quality (i.e. whether adapted spectra appear to have been drawn from the same data distribution as the high SNR spectra used for training). Furthermore, we have shown that our unsupervised stopping criteria was effective for both unsupervised denoising schemes, and appears to be a robust proxy measure of spectral quality (at least in the case here where there is significant domain overlap between unpaired examples). It therefore has the potential for use in other unsupervised denoising schemes. Because the cycleGAN only requires unpaired examples, this approach may ultimately reduce the effort required to leverage the benefits network-based approaches can have on increasing experimental throughput.

\section*{Code and data availability}
The code used for parsing the datasets, training the networks, executing each classical denoising technique, and evaluating the results can be found in \cite{bench_repo}. The parsed datasets used in this study, along with model weights can be found in \cite{bench_drive}.

\section*{Acknowledgments}
This work has received funding from the European Research Council (ERC) under the European Union’s Horizon 2020 research and innovation programme (grant agreement No. 802778). The authors would like to thank Benjamin Gardner (Physics and Astronomy, University of Exeter) for helpful discussions.

%Bibliography
\bibliographystyle{unsrt}  
\bibliography{references}  
\appendix
\section{Additional training details}
\subsection{cycleGAN}
\label{sec:cycleGAN_train}

Here, we elaborate on the procedure for training the denoising cycleGAN.
\begin{enumerate}
    \item A batch of spectra (of size 5) $l$ and $h$ from the low SNR dataset $L$ and the high SNR dataset $H$ are parsed.
    \item Adapted spectra $G_H(l)$ and $G_L(h)$ are generated from the batches.
    \item $G_L(G_H(l))$ and $G_H(G_L(h))$ are computed for the evaluation of the cycle consistency loss.
    \item $G_H(h)$ and $G_L(l)$ are computed for the evaluation of the identity loss.
    \item The adapted and unadapted spectra are fed to their respective discriminators: $D_H(G_H(l))$, $D_L(G_L(h))$, $D_H(h)$, $D_L(l)$. 
    \item{All training loss terms are calculated:}
\begin{itemize}
    \item Generator Loss (adversarial + cycle + identity)
    \begin{itemize}
        \item Adversarial loss (Mean Square Error - MSE)
        \begin{itemize}
            \item MSE$({B_1,D_H(G_H(l))})$
        \item MSE$({B_1,D_L(G_L(h))})$
        \item($B_1$ is a matrix of ones, with the same dimensions as the outputs of $D_L$  and $D_H$)
        \end{itemize}
        \item Cycle (Mean Absolute Error - MAE)
        \begin{itemize}
            \item $\lambda_{C}$MAE$(l,G_L(G_H(l)))$
            \item $\lambda_{C}$MAE$(h,G_H(G_L(h)))$
            \item where $\lambda_{C}=10$
        \end{itemize}
        \item Identity (Mean Absolute Error - MAE)
        \begin{itemize}
            \item $\lambda_{ID}$MAE$(l,G_L(l))$
            \item $\lambda_{ID}$MAE$(h,G_H(h))$
            \item where $\lambda_{ID}=0.5$
        \end{itemize}
    \end{itemize}
    \item Discriminator Losses
        \begin{itemize}
            \item $0.5({\mbox{MSE}(B_{1}, D_L(l)) + \mbox{MSE}(B_{0}, D_L(G_L(h)))})$ 
            \item $0.5({\mbox{MSE}(B_{1}, D_H(h)) + \mbox{MSE}(B_{0}, D_H(G_H(l)))})$ 
            \item $B_{1}$ is a matrix of ones, with the same dimensions as outputs of $D_L$ and $D_H$ 
            \item $B_{0}$ is a matrix of zeros, with the same dimensions as outputs of $D_L$ and $D_H$ 
        \end{itemize}
    
\end{itemize}
\item Then, the generators and discriminators are updated:
\begin{itemize}
    \item Adam used as the optimizer for all networks with a learning rate of $0.0002$, an exponential decay rate for the 1st moment estimates of $\beta_{1} = 0.5$, and the exponential decay rate for the 2nd moment estimates of $\beta_{2}=0.999$. 
\end{itemize}
\end{enumerate}

\begin{table}[H]
   \caption{Hyperparameters used for training the denoising cycleGAN. $\beta_1$ and $\beta_2$ are the exponential decay rates for the first and second moment estimates respectively.}

  \centering
  \begin{tabular}{lll}
    \toprule
    \multicolumn{2}{c}{CycleGAN Hyperparameters}                   \\
    \cmidrule(r){1-2}
    Name     & Setting      \\
    \midrule
    Batch size & 5      \\
    Optimizer & Adam \\
    Learning rates ($G_H, G_L, D_H, D_L$)      & 2e-5   \\
    $\beta_{1}$  ($G_H, G_L, D_H, D_L$)   & .5       \\
    $\beta_{2}$  ($G_H, G_L, D_H, D_L$)   & .999       \\
    \bottomrule
  \end{tabular}
  \label{tab:hyperparams}
\end{table}

\begin{figure}[H]
    \centering
    \includegraphics[width=1\columnwidth]{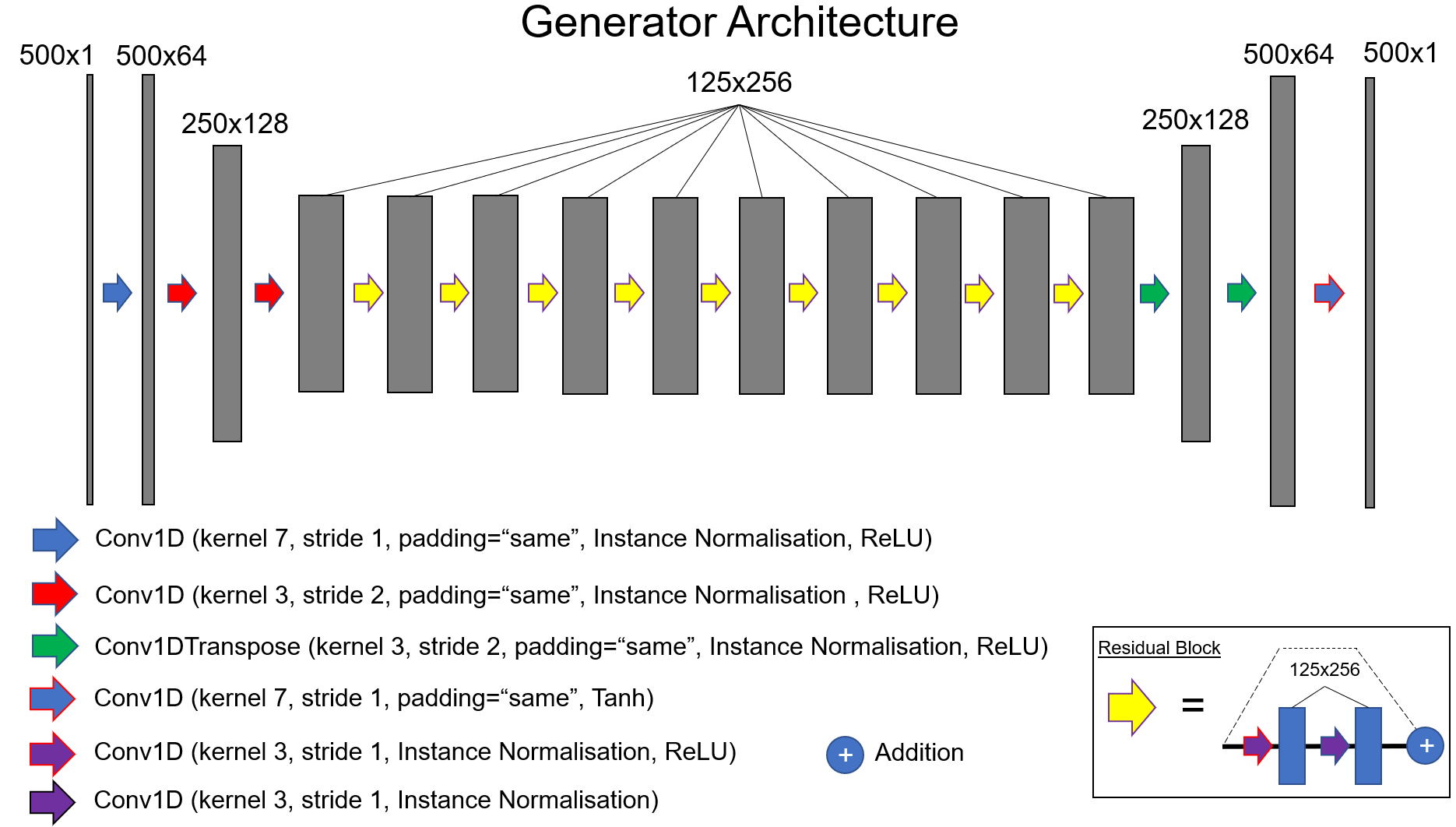}
    \includegraphics[width=.77\columnwidth]{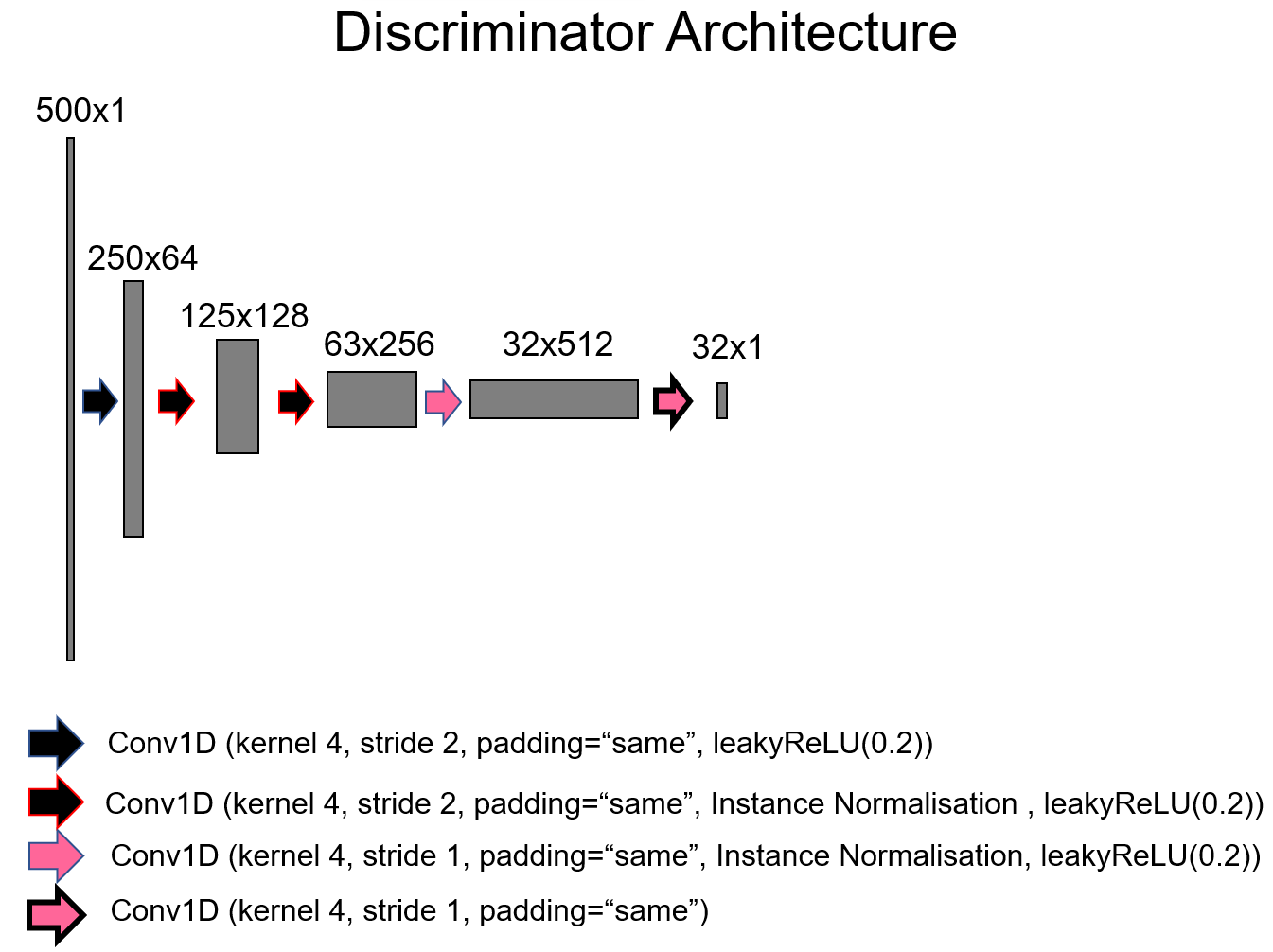}
    \caption{The architectures of the denoising cycleGAN. The activation maps produced by each layer are represented by grey blocks, with their length and the number of activation maps produced by each layer written above them (note that these blocks are not exactly to scale).}
    \label{fig:cycleGAN_archs}
\end{figure}

\subsection{Autoencoder}
\label{sec:AE_details}
Here, we provide additional information about the denoising autoencoder.
\begin{table}[H]
   \caption{Hyperparameters used for training the denoising autoencoder. $\beta_1$ and $\beta_2$ are the exponential decay rates for the first and second moment estimates respectively.}

  \centering
  \begin{tabular}{lll}
    \toprule
    \multicolumn{2}{c}{Autoencoder Hyperparameters}                   \\
    \cmidrule(r){1-2}
    Name     & Setting      \\
    \midrule
    Batch size & 5      \\
    Optimizer & Adam \\
    Learning rate      & 1e-4   \\
    $\beta_{1}$    & .5       \\
    $\beta_{2}$    & .999       \\
    \bottomrule
  \end{tabular}
  \label{tab:AE_hyperparams}
\end{table}

\begin{figure}
    \centering
    \includegraphics[width=.65\columnwidth]{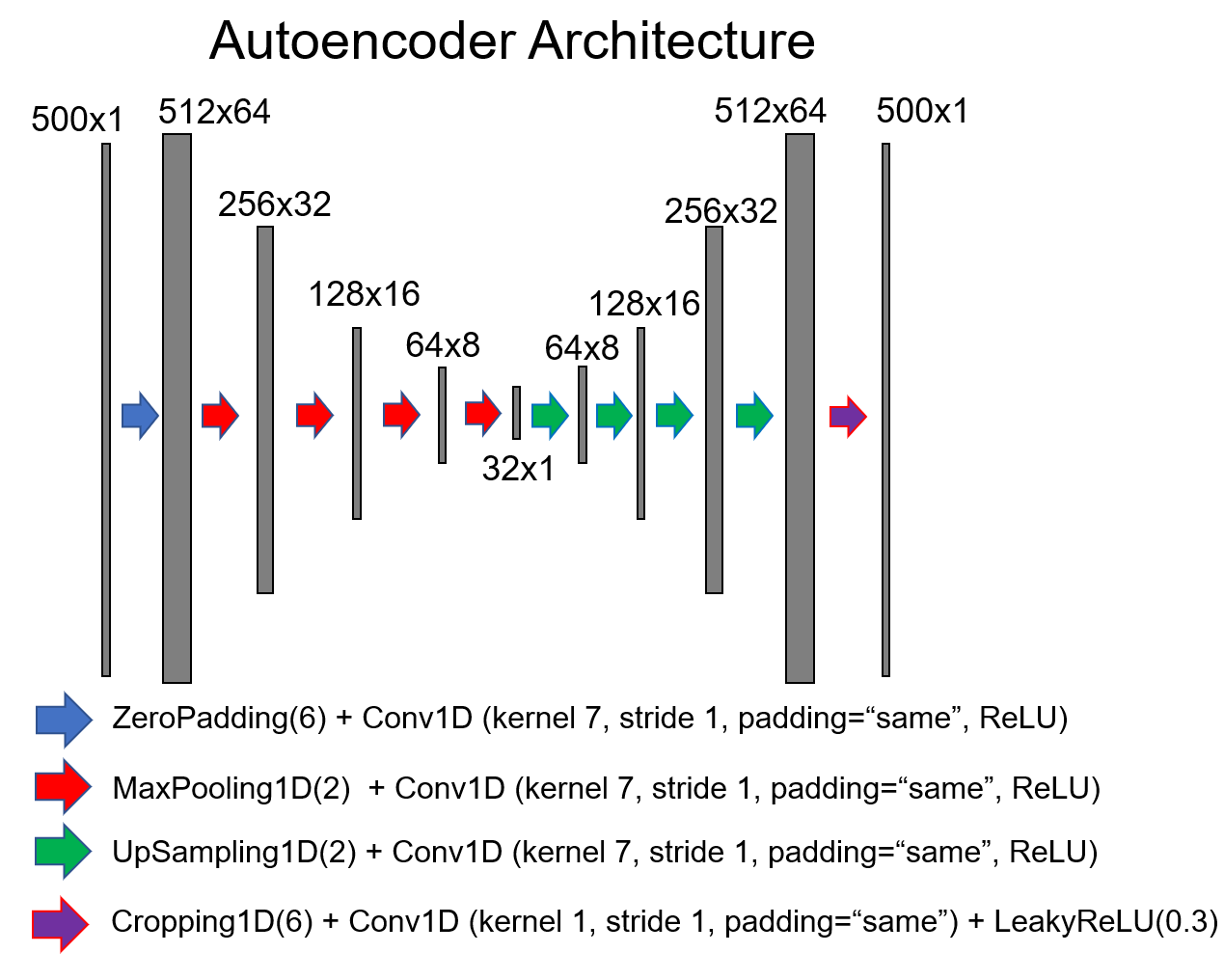}
    \caption{The architecture of the denoising autoencoder. The activation maps produced by each layer are represented by grey blocks, with their length and the number of activation maps produced by each layer written above them (note that these blocks are not exactly to scale).}
    \label{fig:AE_arch}
\end{figure}
\newpage
\section{Clustering details}
\label{sec:cluster_details}
Here, we show the cluster centres used to infer the classes of spectra for calculating the MCSE, and the unsupervised validation loss. We also provide the population of each cluster for the test and validation sets. 
\begin{figure}[H]
    \centering
    \includegraphics[width=.7\columnwidth]{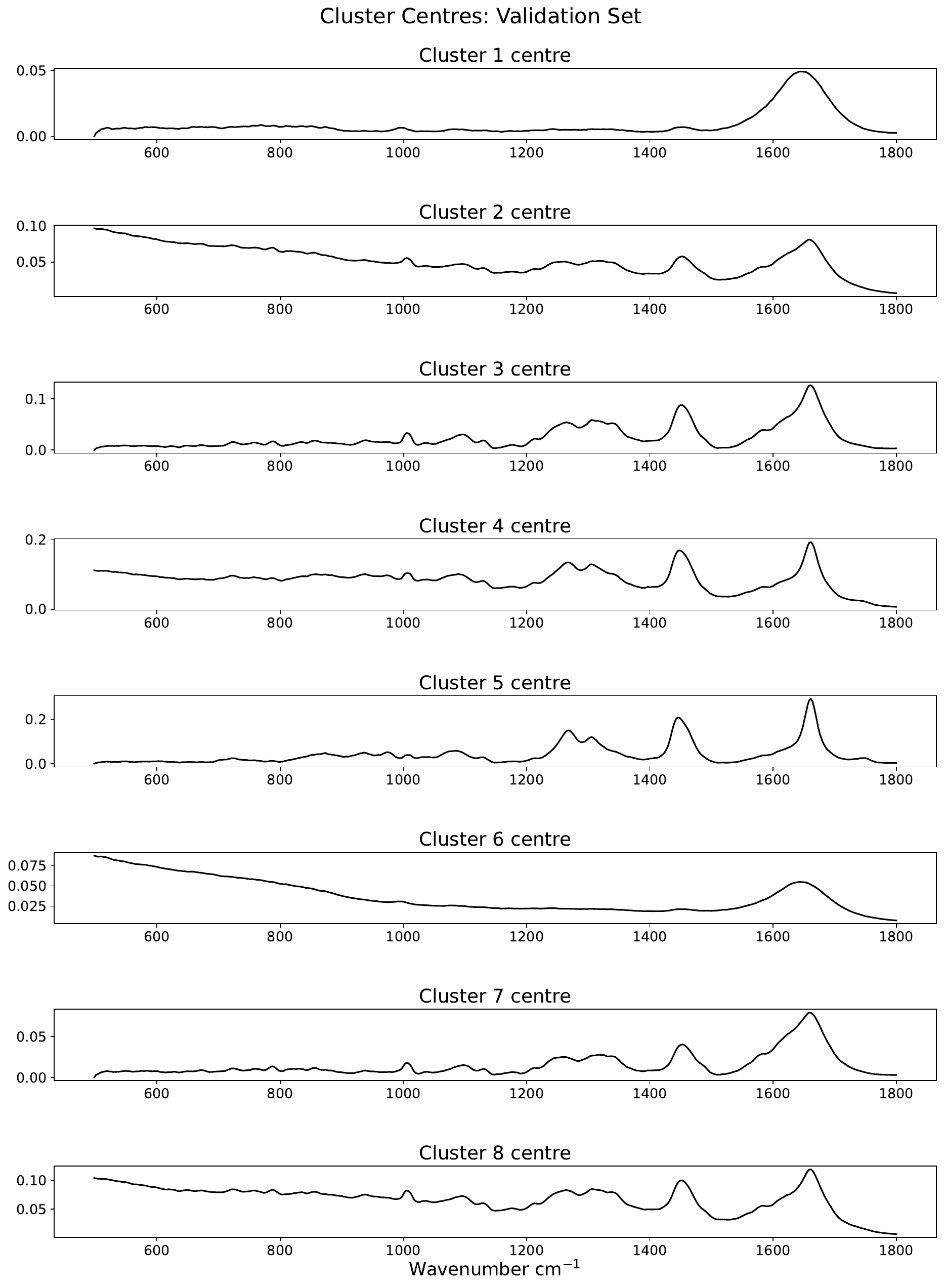}
    \caption{Cluster centres for the cluster high SNR portion of the unsupervised validation set (clustering used to calculate the unsupervised stopping metric).}
    \label{fig:clust_valid}
\end{figure}

\begin{table}[H]
 
  \centering
  \begin{tabular}{lll}
    \toprule
    \multicolumn{2}{c}{Cluster populations validation set}                   \\
    \cmidrule(r){1-2}
    Cluster     & Population      \\
    \midrule
    1 & 4863   \\
    2 & 1682  \\
    3      &1173  \\
    4 &311 \\
    5 &107 & \\
    6 &4821 & \\
    7 &1836 & \\
    8 &1169 & \\
    \bottomrule
  \end{tabular}
  \caption{Cluster populations for the validation set}
  \label{tab:clust_valid}
\end{table}

\begin{figure}[H]
    \centering
    \includegraphics[width=.7\columnwidth]{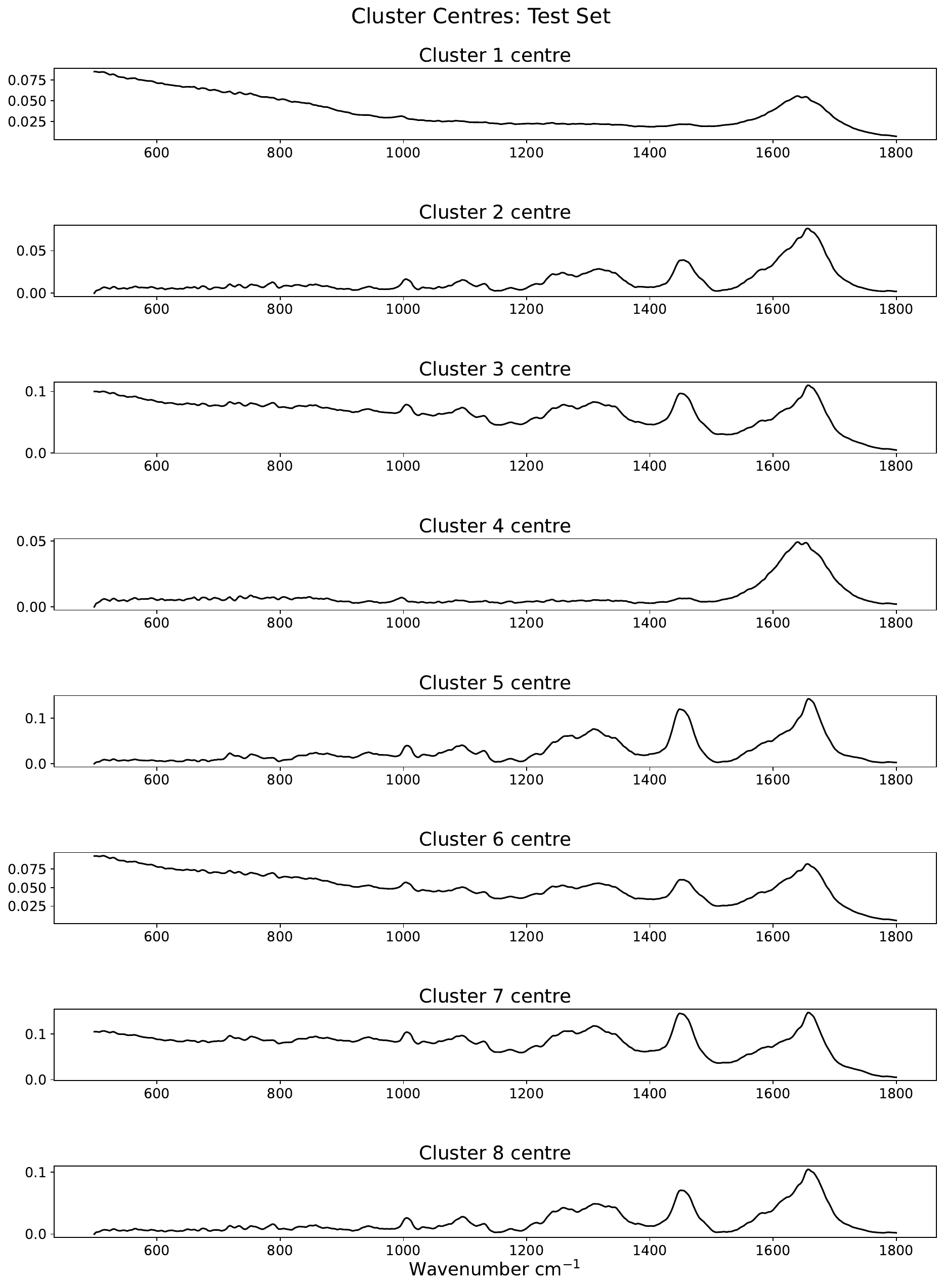}
    \caption{Cluster centres for the test set ground truths (clustering used to calculate the MCSE).}
    \label{fig:clust_test}
\end{figure}

\begin{table}[H]
  \centering
  \begin{tabular}{lll}
    \toprule
    \multicolumn{2}{c}{Cluster populations test set}                   \\
    \cmidrule(r){1-2}
    Cluster     & Population      \\
    \midrule
    1 & 3045   \\
    2 & 1276  \\
    3      &1435  \\
    4 & 3057\\
    5 &628 & \\
    6 &1357 & \\
    7 &510 & \\
    8 &1386 & \\
    \bottomrule
  \end{tabular}
  \caption{Cluster populations for the test set}
  \label{tab:clust_test}
\end{table}
\begin{figure}[H]
    \centering
    \includegraphics[width=.5\columnwidth]{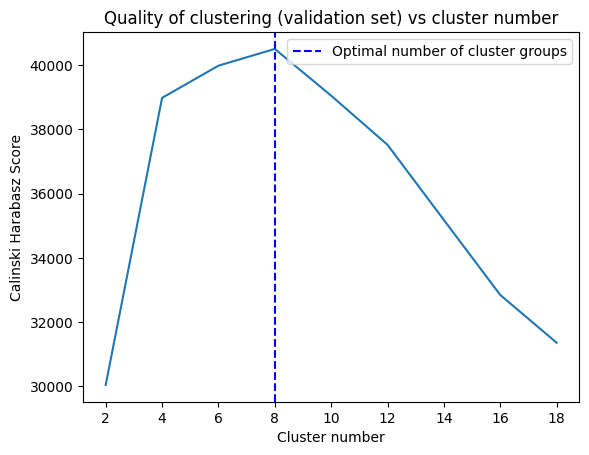}
    \caption{Plot showing the quality of the clustering of the high SNR spectra in the unsupervised validation set (as determined by the Calinksi-Harabasz score) vs the number of cluster groups. Eight groups produced the highest cluster quality, and this was used to compute the unsupervised validation loss.}
    \label{fig:clust_quality_valid_set}
\end{figure}
\newpage
\section{HSI Evaluation}
\label{sec:extended_HSI}
Here, we provide additional slices of the test HSI shown in  Fig. \ref{fig:eval_HSI}.
\begin{figure}[H]
    \centering
    \includegraphics[width=\columnwidth]{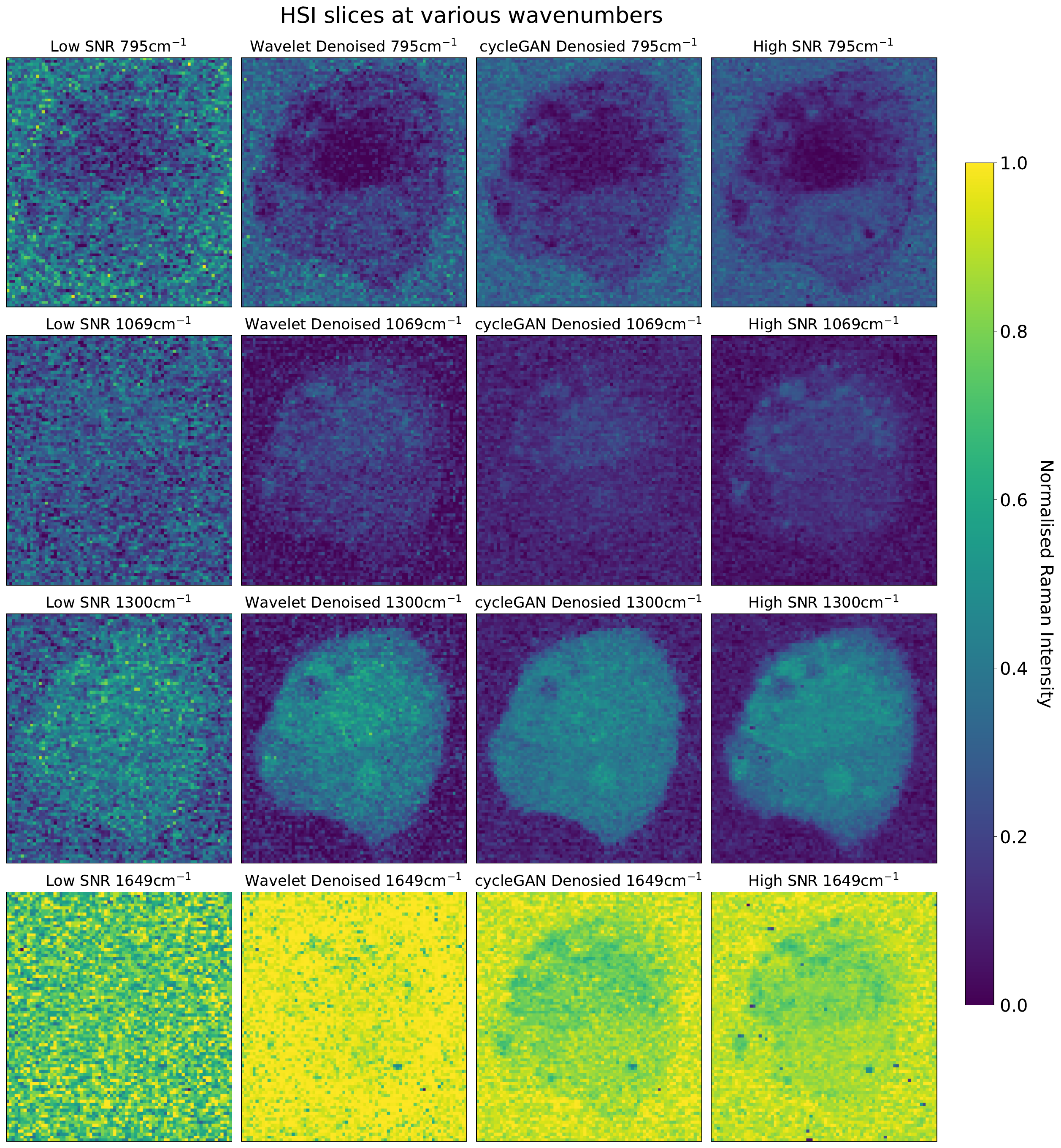}
    \caption{Several slices of the test HSI shown in Fig. \ref{fig:eval_HSI} before and after denoising as well as the high SNR ground truth. Each spectrum was normalised independently of the others which accounts for the high amplitude background signals found in the 1649 cm$^{-1}$ images.}
    \label{fig:eval_HSI_extended1}
\end{figure}
\end{document}